\documentclass[preprint2]{aastex6}
\usepackage{amsmath,natbib,graphicx}
\usepackage{epsf}
\input{epsf}
\usepackage{epstopdf}
\usepackage{multirow}
\usepackage{subfigure}
\usepackage{epsfig}
\usepackage{color}
\usepackage{latexsym}
\usepackage{appendix}
\usepackage{threeparttable}
\usepackage{ifthen}
\usepackage{hyperref}
\usepackage{sistyle}
\SIthousandsep{,}

\DeclareGraphicsExtensions{.jpg,.pdf,.png,.eps,.ps}
\graphicspath{{FIGURES/}}

\newcommand{\beq}{\begin{equation}}
\newcommand{\eeq}{\end{equation}}
\newcommand{\simleq}{{\raise.0ex\hbox{$\mathchar"013C$}\mkern-14mu \lower1.2ex\hbox{$\mathchar"0218$}}}
\newcommand{\simgeq}{{\raise.0ex\hbox{$\mathchar"013E$}\mkern-14mu \lower1.2ex\hbox{$\mathchar"0218$}}}

\newcommand{\wmap}{\textit{WMAP}} 
\newcommand{\planck}{{\sl Planck}}
\newcommand{\herschel}{{\sl Herschel}}

\newcommand{\bk}{\ensuremath{\mathbf{k}}}

\begin{document}

\title{Maps of the Magellanic Clouds from Combined South Pole Telescope and \textit{PLANCK} Data}

\shorttitle{SPT-\planck\ maps of the Magellanic Clouds}
\shortauthors{Crawford et al.}

\def\KICPChicago{1}
\def\AAUChicago{2}
\def\McGill{3}
\def\UChicago{4}
\def\FNAL{5}
\def\ArgonneHEP{6}
\def\PhysicsUChicago{7}
\def\EFIChicago{8}
\def\SLAC{9}
\def\Berkeley{10}
\def\Caltech{11}
\def\MPE{12}
\def\Colorado{13}
\def\LBNL{14}
\def\KIPAC{15}
\def\Davis{16}
\def\Arizona{17}
\def\Michigan{18}
\def\Munich{19}
\def\ExcellenceCluster{20}
\def\Dunlap{21}
\def\Minnesota{22}
\def\Melbourne{23}
\def\CaseWestern{24}
\def\ArtInstChicago{25}
\def\JPL{26}
\def\CfA{27}
\def\Stanford{28}
\def\UToronto{29}
\def\Illast{30}
\def\Illphys{31}

\author{
  T.~M.~Crawford\altaffilmark{\KICPChicago,\AAUChicago},
  R.~Chown\altaffilmark{\McGill},
  G.~P.~Holder\altaffilmark{\McGill},
  K.~A.~Aird\altaffilmark{\UChicago},
  B.~A.~Benson\altaffilmark{\FNAL,\KICPChicago,\AAUChicago},
  L.~E.~Bleem\altaffilmark{\ArgonneHEP,\KICPChicago},
  J.~E.~Carlstrom\altaffilmark{\KICPChicago,\PhysicsUChicago,\ArgonneHEP,\AAUChicago,\EFIChicago},
  C.~L.~Chang\altaffilmark{\ArgonneHEP,\KICPChicago,\AAUChicago},
  H-M.~Cho\altaffilmark{\SLAC},
  A.~T.~Crites\altaffilmark{\KICPChicago,\AAUChicago,\Caltech},
  T.~de~Haan\altaffilmark{\McGill,\Berkeley},
  M.~A.~Dobbs\altaffilmark{\McGill},
  E.~M.~George\altaffilmark{\Berkeley,\MPE},
  N.~W.~Halverson\altaffilmark{\Colorado},
  N.~L.~Harrington\altaffilmark{\Berkeley},
  W.~L.~Holzapfel\altaffilmark{\Berkeley},
  Z.~Hou\altaffilmark{\KICPChicago,\AAUChicago},
  J.~D.~Hrubes\altaffilmark{\UChicago},
  R.~Keisler\altaffilmark{\KICPChicago,\PhysicsUChicago,\KIPAC},
  L.~Knox\altaffilmark{\Davis},
  A.~T.~Lee\altaffilmark{\Berkeley,\LBNL},
  E.~M.~Leitch\altaffilmark{\KICPChicago,\AAUChicago},
  D.~Luong-Van\altaffilmark{\UChicago},
  D.~P.~Marrone\altaffilmark{\Arizona},
  J.~J.~McMahon\altaffilmark{\Michigan},
  S.~S.~Meyer\altaffilmark{\KICPChicago,\AAUChicago,\EFIChicago,\PhysicsUChicago},
  L.~M.~Mocanu\altaffilmark{\KICPChicago,\AAUChicago},
  J.~J.~Mohr\altaffilmark{\Munich,\ExcellenceCluster,\MPE},
  T.~Natoli\altaffilmark{\KICPChicago,\PhysicsUChicago,\Dunlap},
  S.~Padin\altaffilmark{\KICPChicago,\AAUChicago},
  C.~Pryke\altaffilmark{\Minnesota},
  C.~L.~Reichardt\altaffilmark{\Berkeley,\Melbourne},
  J.~E.~Ruhl\altaffilmark{\CaseWestern},
  J.~T.~Sayre\altaffilmark{\CaseWestern,\Colorado},
  K.~K.~Schaffer\altaffilmark{\KICPChicago,\EFIChicago,\ArtInstChicago},
  E.~Shirokoff\altaffilmark{\Berkeley,\KICPChicago,\AAUChicago}, 
  Z.~Staniszewski\altaffilmark{\CaseWestern,\JPL},
  A.~A.~Stark\altaffilmark{\CfA},
  K.~T.~Story\altaffilmark{\KICPChicago,\PhysicsUChicago,\KIPAC,\Stanford},
  K.~Vanderlinde\altaffilmark{\McGill,\Dunlap,\UToronto},
  J.~D.~Vieira\altaffilmark{\Illast,\Illphys}, and
  R.~Williamson\altaffilmark{\KICPChicago,\AAUChicago}
  }

\altaffiltext{\KICPChicago}{Kavli Institute for Cosmological Physics, University of Chicago, Chicago, IL, USA 60637}
\altaffiltext{\AAUChicago}{Department of Astronomy and Astrophysics, University of Chicago, Chicago, IL, USA 60637}
\altaffiltext{\McGill}{Department of Physics, McGill University, Montreal, Quebec H3A 2T8, Canada}
\altaffiltext{\UChicago}{University of Chicago, Chicago, IL, USA 60637}
\altaffiltext{\FNAL}{Fermi National Accelerator Laboratory, MS209, P.O. Box 500, Batavia, IL 60510}
\altaffiltext{\ArgonneHEP}{High Energy Physics Division, Argonne National Laboratory, Argonne, IL, USA 60439}
\altaffiltext{\PhysicsUChicago}{Department of Physics, University of Chicago, Chicago, IL, USA 60637}
\altaffiltext{\EFIChicago}{Enrico Fermi Institute, University of Chicago, Chicago, IL, USA 60637}
\altaffiltext{\SLAC}{SLAC National Accelerator Laboratory, 2575 Sand Hill Road, Menlo Park, CA 94025}
\altaffiltext{\Berkeley}{Department of Physics, University of California, Berkeley, CA, USA 94720}
\altaffiltext{\Caltech}{California Institute of Technology, Pasadena, CA, USA 91125}
\altaffiltext{\MPE}{Max-Planck-Institut f\"{u}r extraterrestrische Physik, 85748 Garching, Germany}
\altaffiltext{\Colorado}{Department of Astrophysical and Planetary Sciences and Department of Physics, University of Colorado, Boulder, CO, USA 80309}
\altaffiltext{\LBNL}{Physics Division, Lawrence Berkeley National Laboratory, Berkeley, CA, USA 94720}
\altaffiltext{\KIPAC}{Kavli Institute for Particle Astrophysics and Cosmology, Stanford University, 452 Lomita Mall, Stanford, CA 94305}
\altaffiltext{\Davis}{Department of Physics, University of California, Davis, CA, USA 95616}
\altaffiltext{\Arizona}{Steward Observatory, University of Arizona, 933 North Cherry Avenue, Tucson, AZ 85721}
\altaffiltext{\Michigan}{Department of Physics, University of Michigan, Ann  Arbor, MI, USA 48109}
\altaffiltext{\Munich}{Faculty of Physics, Ludwig-Maximilians-Universit\"{a}t, 81679 M\"{u}nchen, Germany}
\altaffiltext{\ExcellenceCluster}{Excellence Cluster Universe, 85748 Garching, Germany}
\altaffiltext{\Dunlap}{Dunlap Institute for Astronomy \& Astrophysics, University of Toronto, 50 St George St, Toronto, ON, M5S 3H4, Canada}
\altaffiltext{\Minnesota}{Department of Physics, University of Minnesota, Minneapolis, MN, USA 55455}
\altaffiltext{\Melbourne}{School of Physics, University of Melbourne, Parkville, VIC 3010, Australia}
\altaffiltext{\CaseWestern}{Physics Department, Center for Education and Research in Cosmology and Astrophysics, Case Western Reserve University,Cleveland, OH, USA 44106}
\altaffiltext{\ArtInstChicago}{Liberal Arts Department, School of the Art Institute of Chicago, Chicago, IL, USA 60603}
\altaffiltext{\JPL}{Jet Propulsion Laboratory, California Institute of Technology, Pasadena, CA 91109, USA}
\altaffiltext{\CfA}{Harvard-Smithsonian Center for Astrophysics, Cambridge, MA, USA 02138}
\altaffiltext{\Stanford}{Dept. of Physics, Stanford University, 382 Via Pueblo Mall, Stanford, CA 94305}
\altaffiltext{\UToronto}{Department of Astronomy \& Astrophysics, University of Toronto, 50 St George St, Toronto, ON, M5S 3H4, Canada}
\altaffiltext{\Illast}{Astronomy Department, University of Illinois at Urbana-Champaign, 1002 W. Green Street, Urbana, IL 61801, USA}
\altaffiltext{\Illphys}{Department of Physics, University of Illinois Urbana-Champaign, 1110 W. Green Street, Urbana, IL 61801, USA}

\email{tcrawfor@kicp.uchicago.edu}

\slugcomment{Published in The Astrophysical Journal Supplement: 227 (2016) 23}

\begin{abstract}

We present maps of the Large and Small Magellanic Clouds from combined South Pole Telescope (SPT) and \planck\ data. The \planck\ satellite observes in  nine bands, while the SPT data used in this work were taken with the three-band SPT-SZ camera, The SPT-SZ bands correspond closely to three of the nine \planck\ bands,  namely those centered at 1.4, 2.1, and 3.0 mm. The angular resolution of the \planck\ data ranges from 5 to 10 arcmin, while the SPT resolution ranges from 1.0 to 1.7 arcmin. The combined maps take advantage of  the high resolution of the SPT data and the long-timescale stability of the space-based \planck\ observations to  deliver robust brightness measurements on scales from the size of the maps down to $\sim$1 arcmin. In each band, we first calibrate and color-correct the SPT data to match the \planck\ data, then we use noise estimates from each instrument and knowledge of each instrument's beam to make the inverse-variance-weighted combination of the two instruments' data as a function of angular scale. We create maps assuming a range of underlying emission spectra and at a range of final resolutions. We perform several consistency tests on the combined maps and estimate the expected noise in measurements of features in the maps. We compare maps from this work to maps from the \herschel\  HERITAGE survey, finding general consistency between the datasets. All data products described in this paper are available for download from the NASA Legacy Archive for Microwave Background Data Analysis server.

\end{abstract}

\keywords{(galaxies:) Magellanic Clouds --- methods: data analysis}

\section{Introduction}
\label{sec:intro}

The dwarf galaxies known as the Large and Small Magellanic Clouds (LMC and SMC) 
are the most easily observable
extragalactic features in the sky and have been the subject of hundreds of years of observation
(see, e.g., \citealt{westerlund97} for a review). Among the most active areas of research involving 
the Magellanic Clouds is their use as laboratories in which to study star formation. Several features of the 
LMC and SMC make them particularly useful for studies of star formation, including their proximity
(at $\sim 50$ and $\sim 60$~kpc, respectively, they are the nearest high-contrast extragalactic systems), their 
orientation (we see the LMC nearly face-on), and the diversity in key star-formation observables 
(such as metallicity, gas density, and gas-to-dust ratio) among the LMC, SMC, and Milky Way 
\citep{mizuno09,meixner13}. Furthermore, the distances to the Magellanic Clouds are well-determined,
unlike distances to many features in the Milky Way, so absolute luminosities of features in the 
LMC and SMC can be determined with fairly high precision.

Continuum observations in the far-infrared (FIR), submillimeter (submm), and millimeter (mm) bands
can provide important constraints on star formation scenarios through the sensitivity of such bands
to thermal dust emission, as well as free-free and synchrotron emission from active regions
(e.g., \citealt{dezotti10}, \citealt{boselli11}). Until roughly a decade ago, there were relatively few
robust measurements of the Magellanic Clouds at these wavelengths, particularly in the mm and 
submm bands. The launch of the 
\wmap,\footnote{\url{http://map.gsfc.nasa.gov}}
\planck,\footnote{\url{http://www.cosmos.esa.int/web/planck}}
and 
\herschel \footnote{\url{http://www.cosmos.esa.int/web/herschel}}
satellites fundamentally changed
this situation. Using data from the balloon-borne TopHat instrument \citep{aguirre03} and the \wmap\ 
satellite, \citet{israel10} noted a significant excess in mm/submm emission 
(relative to the modified blackbody models usually assumed to describe thermal dust emission)
in the Magellanic Clouds, 
particularly the SMC. These results were confirmed with data from the \planck\ satellite
\citep{planck11-17} at lower noise and higher resolution (roughly 5~arcmin in the shortest-wavelength
\planck\ bands). More recently, the HERITAGE survey using the \herschel\ satellite 
\citep{meixner13} has produced
sub-arcminute-resolution maps of the LMC and SMC in five bands spanning wavelengths from 
100 to 500~$\mu$m.

The aim of this paper is to extend the wavelength range of arcminute-resolution maps of the LMC and
SMC by combining \planck\ data with data from the 10-meter South Pole Telescope (SPT, \citealt{carlstrom11}).
The SPT is a ground-based telescope that has so far been configured to observe
in up to three mm bands, each of which has a counterpart of similar central wavelength and bandwidth
among the \planck\ observing bands.
The combination of instantaneous sensitivity 
and resolution of the SPT is nearly unparalleled in these bands, but it is difficult to measure emission
at very large scales (degree-scale and larger) from the ground because of atmospheric contamination. 
To obtain an unbiased estimate of the brightness of the LMC and SMC across the full range of angular 
scales---from the arcminute SPT beam to the many-degree extent of these galaxies (roughly $7^\circ$ 
for the LMC)---in this work we combine the small-scale information from SPT with the larger-scale
information from the corresponding bands in \planck\ satellite data. 
The primary science goal of both SPT and \planck\ is to measure temperature and polarization 
anisotropy in the cosmic microwave background (CMB), and similarly combined maps of low-emission
regions of the sky will be useful for cosmological studies. In one sense, this work is a pilot project
for these future studies; however, we expect the data products that result from this analysis will be 
immediately useful to a wide range of astronomical applications.

This paper is structured as follows. In Section~\ref{sec:data}, we describe the SPT and \planck\ 
instruments and data products. In Section~\ref{sec:combine}, we describe the procedure we use
to combine the two data sets into a single map in each observing band. In Section~\ref{sec:results}, 
we present the combined maps and perform a number of quality-control checks. 
In Section~\ref{sec:herschel}, we compare the combined maps with FIR/submm maps
from the \herschel\ HERITAGE survey. We conclude in
Section~\ref{sec:conclusions}.

\section{Instruments, Data, and Processing}
\label{sec:data}

\subsection{SPT}
\label{sec:spt}
The SPT
is a 10-meter telescope located within
1~km of the geographical South Pole, at the National Science Foundation 
Amundsen-Scott South Pole station. 
The telescope is designed for millimeter and sub-millimeter observations of faint, diffuse sources, 
in particular anisotropy in the CMB.
From 2007 to 2011, the instrument at the focus of the
SPT was the SPT-SZ camera, which consisted of 960 detectors in three wavelength bands centered at 
roughly 1.4, 2.0, and 3.2 mm (center frequencies of roughly 220, 150, and 95~GHz). 
The main lobe of the instrument beam, or point-spread function,
is closely approximated by an azimuthally symmetric, two-dimensional Gaussian.
The main-lobe full width at half maximum (FWHM) 
measured on bright point sources in survey fields (which includes a contribution
from day-to-day pointing variations) is equal 
to 1.0, 1.2, and 1.7~arcmin at 1.4, 2.0, and 3.2~mm, respectively.

\subsubsection{SPT Observations of the Magellanic Clouds}
\label{sec:sptobs}
In 2011 November, parts of three observing days were spent on
dedicated observations of fields centered on the Magellanic Clouds. 
The bulk of the time---roughly 20 hours---was spent on the LMC, with approximately
three hours spent on the SMC. 
The LMC field was defined as an $8^\circ$-by-$8^\circ$
region centered at R.A.~$80^\circ$, declination $-68.5^\circ$.
The SMC field was defined as a $5^\circ$-by-$5^\circ$
region centered at R.A.~$15^\circ$, declination $-72.5^\circ$.
As with most fields observed with the SPT, these
observations were conducted by scanning the telescope back and forth in azimuth
then taking a small (6 arcmin) step in elevation. Because of the geographical location
of the telescope, this corresponds to scanning in right ascension and stepping in 
declination. At the scan speed used for these observations ($\sim 0.4^\circ$/s on the sky),
this scan pattern covers the LMC field in 90 minutes and the SMC field in 45 minutes.
We refer to each individual 90- or 45-minute set of scans as an ``observation.''

\subsubsection{Data Processing}
\label{sec:sptdata}
Detector data are processed into maps individually for each observation and 
wavelength band. The processing pipeline used in this work is described in detail
in \citet{schaffer11}; we summarize it briefly here. For each observation, data that pass 
cuts are flat-fielded (by adjusting the data from each detector according to the 
response of that detector to an internal calibration source) and filtered. Using inverse-variance weighting, 
the data are binned into pixels based on the value of the telescope boresight pointing in 
every data sample and the known physical locations of the detectors in the focal plane. 
The maps for this work are 
made in the oblique Lambert equal-area azimuthal (ZEA) projection, with a pixel
scale of 0.25~arcmin.

\begin{figure}
\begin{centering}
\includegraphics[width=3in]{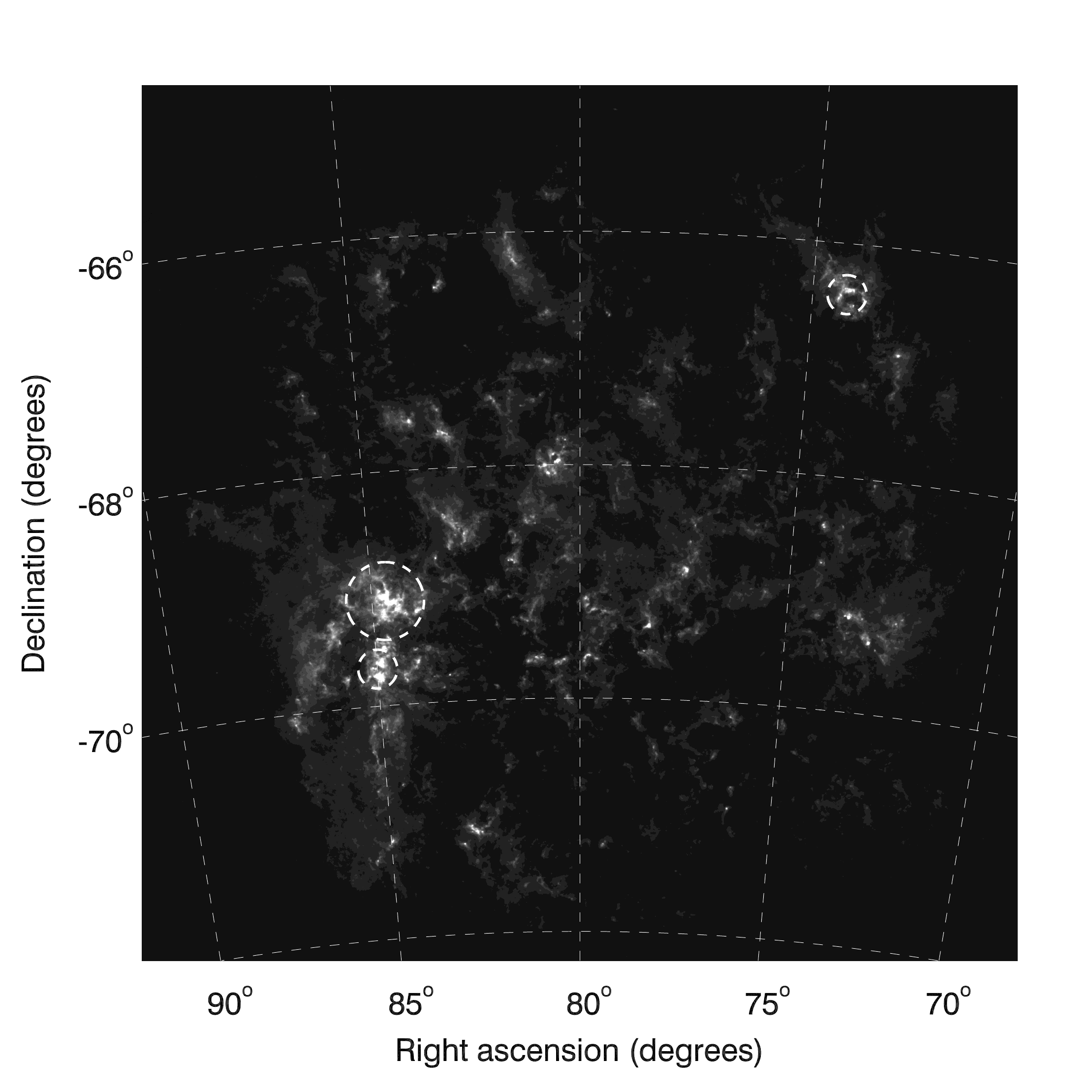}
\includegraphics[width=3in]{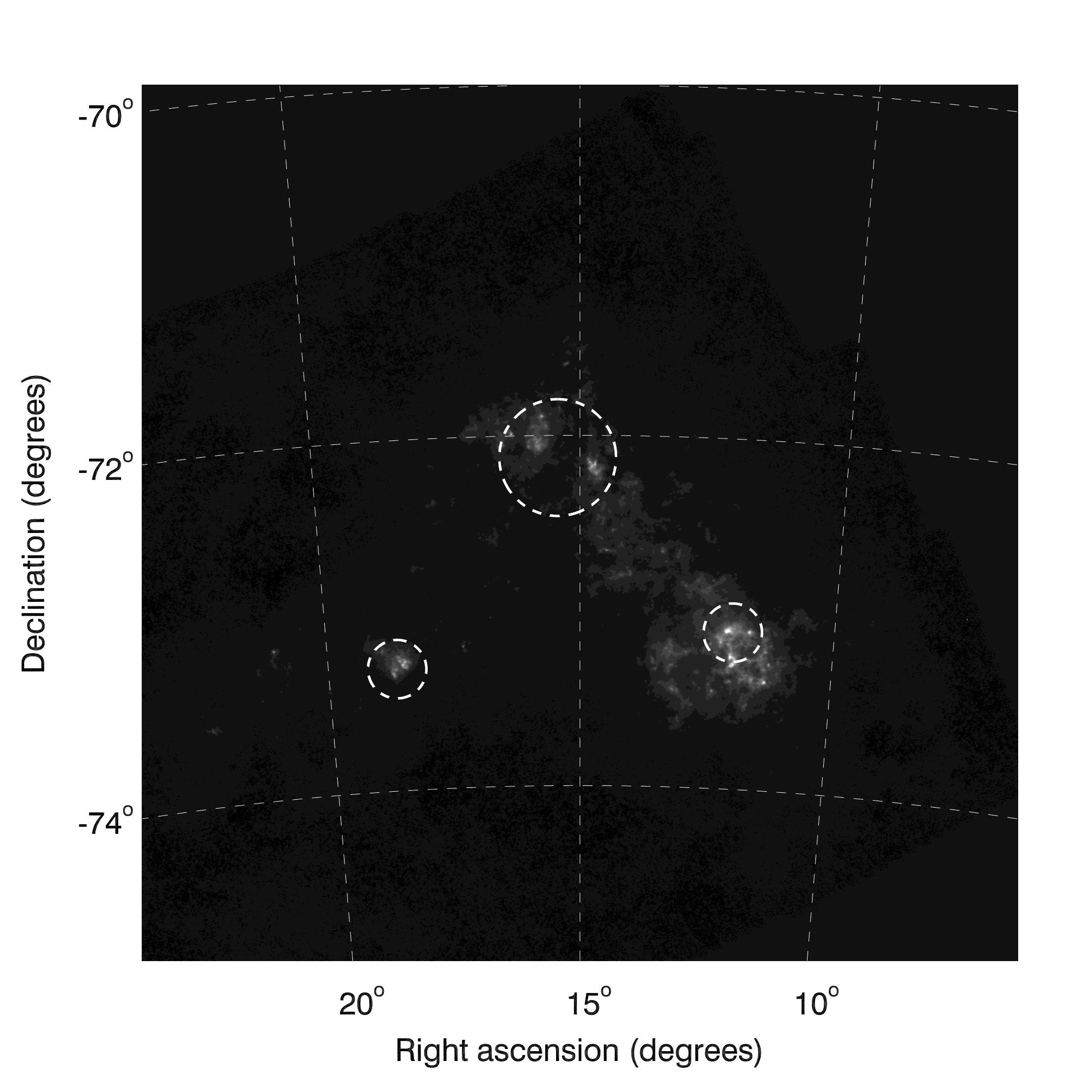}
\caption{
\label{fig:maskreg}
Regions masked in the filtering of SPT time-ordered data. 
\textbf{Top panel}: 500~$\mu$m map of the LMC from the \herschel\ 
HERITAGE survey with the three masked LMC regions indicated by
dashed circles (see Table~\ref{tab:mask} for exact locations).
\textbf{Bottom panel}: 500~$\mu$m map of the SMC from the \herschel\ 
HERITAGE survey with the three masked SMC regions indicated by
dashed circles (see Table~\ref{tab:mask} for exact locations).
}
\end{centering}
\end{figure}

The filtering applied to the data consists of three steps, the first two of which 
are primarily to suppress the effects of atmospheric noise. 
First, a fifth-order polynomial is fit to the data from each detector in each scan 
and then subtracted from that data.
Next, at every time sample, the mean across a detector module (there are six modules 
in the SPT-SZ camera, each with 160 detectors of a given frequency) and two spatial 
gradients across that module are calculated and subtracted from the data of each 
detector on that module. Finally, a Fourier-domain low-pass filter is applied to each 
detector's data to avoid aliasing when the data are binned into map pixels.
In the polynomial subtraction step, certain very bright regions of each field 
are not included in the polynomial fit, in an effort to avoid large filtering artifacts
around these regions that could affect measurements of nearby regions. We mask
three regions in each field. These regions are selected by visually inspecting 2.0~mm
maps made without masking and selecting the regions with the largest filtering
artifacts. The centers and extents of the masked regions are
listed in Table~\ref{tab:mask} and shown on 500~$\mu$m \herschel\ images of the LMC
and SMC in Figure~\ref{fig:maskreg}. These regions are not masked in the module-based
spatial mode subtraction, because the modes down-weighted 
by this filtering are well measured by \planck\ and will be properly represented in the
combined map (see Section \ref{sec:combine}
for details). 

In Section~\ref{sec:checks}, we discuss the slight bias in aperture photometry incurred
by filtering out certain angular modes from the SPT data and not replacing those modes
with \planck\ data. The bias is typically on the order of 2\%. This bias does not affect the 
regions that were masked in filtering.

\begin{deluxetable}{c c c c}
\tablecaption{Regions masked in the SPT time-ordered data polynomial subtraction}
\tablehead{
\colhead{Field} &
\colhead{Mask center R.A.} &
\colhead{Mask center decl.} & 
\colhead{Mask radius} \\
\colhead{   } &
\colhead{[deg.]} &
\colhead{[deg.]} & 
\colhead{[arcmin]} 
}
\startdata
LMC &     84.684  &      -69.105   &    20 \\ 
LMC &     74.265  &      -66.437   &    10 \\
LMC &     84.991  &      -69.682   &    10 \\
SMC &    15.414  &       -72.127   &    20 \\
SMC &    11.995   &     -73.105    &    10 \\
SMC &    18.635  &      -73.304    &    10
\enddata
\label{tab:mask}
\end{deluxetable}

The individual-observation maps in each observing band are combined into full coadded 
maps using inverse-variance weighting. If the data from one observing band in 
one individual observation has too few detectors that pass cuts, or if any obvious
artifacts are seen when the single-observation map is visually inspected, 
the map from that observation
is not included in the coadded map. Of the 14 individual LMC field observations, 10 are
used in the 1.4~mm coadd, 12 in the 2.0~mm coadd, and 12 in the 3.2~mm coadd. Of the
four individual SMC field observations, three are used in the 1.4~mm coadd and all four at 
2.0 and 3.2~mm. The most common reason for detectors failing cuts is poor weather, which
affects the shorter wavelengths more severely (because of the spectral dependence of
atmospheric noise at millimeter wavelengths---see, e.g., \citealt{bussmann05}).

In addition to the coadded signal maps, we create coadded null maps for each observing
band and field. We combine these maps with \planck\ HFI null maps in the same
way as signal maps are combined, such that the combined null maps can be used in estimating
the noise contribution to the uncertainty on any quantity estimated from the combined signal
maps. For SPT, we create null maps by subtracting maps made from data in right-going telescope 
scans only from maps made from data in left-going telescope scans only (divided by two). 
Any true sky signal
should difference away in this operation, leaving an estimate of the instrumental and atmospheric
noise. 
The individual-observation null maps are combined in the same way as the 
individual-observation signal maps, except that an additional layer of differencing is performed
by multiplying one half of the observations by -1. Despite this double differencing
(left minus right, multiplying half the individual-observation maps by -1), some small 
artifacts are visible in the null maps at the
location of the brightest regions of the two fields---most notably at the location of 30 Doradus
in the LMC. These are due to slight differences in weights and filtering in the left-going and 
right-going maps, and the amplitude of the artifacts are at most 1\% of the amplitude of the
original features.

\subsubsection{Angular Response Function}
\label{sec:sptxfer}
As mentioned above, the true instrument beam in each SPT observing band---i.e., the response 
to a point source as a function of angular offset from the source
that would be measured in the absence of any processing to the 
data---is well-approximated by an azimuthally symmetric Gaussian. These beams are estimated
from a combination of dedicated observations of planets and measurements of bright point sources
in the SPT-SZ survey field (for details, see \citealt{schaffer11}).
The effect of the filtering of SPT data is to modify this angular response function---i.e., to alter the 
effective instrument beam. Each filtering step has a specific impact on the effective beam.
The polynomial subtraction imparts slight negative lobes to the beam in the scan direction---in
this case R.A. or $x$---while the module-based filtering imparts an isotropic negative ring at 
roughly half the scale of a module, or $\sim10$~arcmin. The anti-aliasing filter smooths
the data in the scan direction at or just above the pixel scale (0.25~arcmin); this smoothing
is negligible compared to the size of the true beam. All of these effects are 
represented more cleanly in the two-dimensional Fourier domain, and we use Fourier methods to estimate and
represent the response function in this work.

The filter response function is estimated using simulated observations. One hundred independent
simulated skies are created, in which the sky signal is white noise 
convolved with a Gaussian with FWHM equal to 0.75~arcmin.
For each simulated sky, a simulated version of the full time-ordered data 
in each real observation of the LMC or SMC field is created using the telescope pointing and
detector focal plane locations. These simulated time-ordered data are then filtered and made 
into a map in the same manner as is used for the real data, including detector cuts and weighting. 
The individual-observation maps are combined into full coadded maps using the same procedure 
and weighting as for the real data. For each of the 100 simulated skies, the square of the 
two-dimensional Fourier transform of the coadded map is divided by the known input (2d) power 
spectrum. These 100 estimates are averaged, and the square root of the result is our estimate of
the 2d filter response function. We multiply this (in Fourier space) by the instrument beam 
to create the full beam-plus-filtering response function.

\begin{figure}
\includegraphics[width=3.5in]{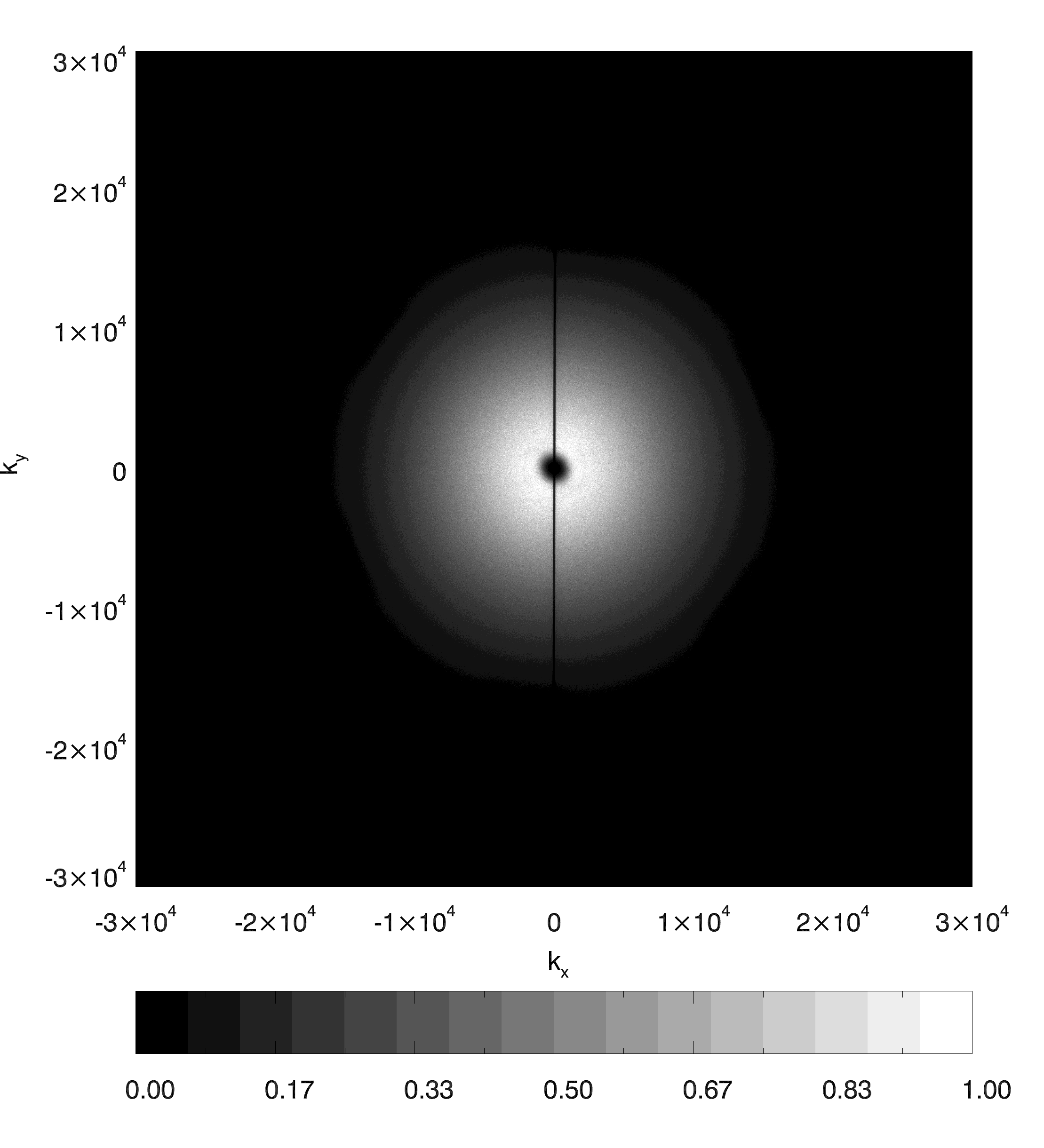}
\caption{Two-dimensional Fourier-domain angular response function for the 2.0~mm SPT data
used in this work.
The response function is the product of the instrument beam or point-spread function and the
filtering performed on the data. The isotropic suppression of power at $k_x \simeq k_y \simeq 0$ is 
from the subtraction of a common mode and two slopes across each detector module at 
each time sample. The thin line of zero power along $k_x=0$ is from the subtraction of a 
fifth-order polynomial from the data from each detector individually on each scan across the field. The isotropic
rolloff at high $k$ is due to the beam.
}
\label{fig:tf150}
\end{figure}

Figure \ref{fig:tf150} shows the full two-dimensional Fourier-domain angular response function 
(beam plus filtering) for the 2.0~mm SPT data used in this work. The filter part of the response functions 
for the 1.4 and 3.2~mm data are nearly identical to the filter part of the 2.0~mm response function.
The effect of each filtering step is confined to a specific region of 2d Fourier space.
The polynomial subtraction acts as a one-dimensional high-pass filter, suppressing modes at $k_x < 100$,
while the module-based filter acts as an isotropic high-pass, suppressing modes at 
$k < 1000$, where $k$ is angular wavenumber ($k(\lambda)=2 \pi/\lambda$ for wavelength $\lambda$
in radians), and $k_x$ is the Fourier conjugate of the scan direction. The anti-aliasing filter acts as
a scan-direction low-pass filter with a cutoff at $k_x \simeq \num{20000}$; however the effect of this low-pass
is dominated by the isotropic low-pass of the instrument beam and is not visible in Figure \ref{fig:tf150}.

Finally, we note that the clean representation of the filter response function in 2d Fourier 
space is to some degree dependent on the projection used to map the curved sky onto
a flat, 2d grid. In particular, any filtering that acts on single-detector time-ordered data 
will result in an effective map-space filter
along the scan direction on the sky. Any projection in which the scan 
direction (R.A.) corresponds to the $x$ axis of the 2d map will localize this filtering 
in 2d Fourier space to a particular region of 1d angular frequency or wavenumber $k_x$.
This makes it easy to identify which Fourier modes in the map have been downweighted 
by the filtering and to replace those modes with modes from the corresponding \planck\ HFI
map. The downside of such a projection is that the mapping of R.A. to $x$ everywhere in 
the map necessarily leads to angular distortions at the map edges. Such a projection is 
not optimal for representing the true instrument beam in 2d Fourier
space; an angle-preserving projection such as the ZEA projection
is more appropriate for dealing with the beam. 
The maps used to create the representation of the filter function in Figure~\ref{fig:tf150} were
made in a simple Cartesian projection, but all other maps in this analysis are made in
the ZEA projection.

\subsubsection{Noise Estimation}
\label{sec:sptpsd}
To combine SPT data with \planck\ HFI data in a nearly optimal way, we need a measure of not
only the angular response function for each instrument but also a measure of the noise in each
data set. For SPT, the noise is most cleanly represented in the two-dimensional Fourier domain
(as was the case with the SPT angular response function). We estimate the 2d noise power spectrum by coadding 
the single-observation maps with half the maps multiplied by -1, taking the Fourier transform of 
the result and squaring, and repeating many times with the negative sign assigned to a different 
set of single-observation maps each time. For the SMC field, there are not enough single-observation
maps to get a good noise estimate with this technique, so we use the LMC field estimate scaled
by the ratio of observing depth for our SMC field estimate. We note that using a single 2d Fourier
estimate for the noise over an entire field assumes that the noise properties are uniform over 
the field. This is a very good approximation for the SPT maps in this work, except for small regions
at the edges of the fields which are not used in the combination with \planck\ HFI data.

\subsubsection{Filter Deconvolution}
\label{sec:sptdecon}
In preparation for combining the SPT maps with \planck\ HFI maps, we deconvolve the filter
angular response function from the maps, and we modify the noise estimates to account for 
this deconvolution. We perform this deconvolution in 2d Fourier space: after first multiplying the map
by a real-space apodization window, we Fourier transform the map, multiply in Fourier space by 
the reciprocal of the
filter response function, and inverse Fourier transform. To avoid numerical 
issues, we set the reciprocal of the filter response to zero in any region of 2d Fourier space 
in which the filter response is less than 0.01. This conditioning step is taken into account when 
we combine the SPT and \planck\ maps. We account for the deconvolution in the SPT noise
estimates by multiplying the 2d Fourier-space noise estimates by the (conditioned) reciprocal 
of the filter response.

Note that we only deconvolve the azimuthally symmetric, low-$k$ part of the 
filter response (the part of Fourier space in which the data can be adequately replaced
with \planck\ data)
while leaving the low-$k_x$, high-$k_y$ part of the response function in the map. This 
means that in the final, combined maps, a small fraction of angular modes will be missing
from the data (except in the regions which were masked during this filtering step).
As can be seen from Figures~\ref{fig:tf150} and \ref{fig:noisevscale}, after 
combining with \planck\ data, modes will be missing from a small area at $k_x \lesssim 100$
and $k_y \gtrsim 2000$.
See Section \ref{sec:checks} for a discussion of the effects of ignoring this small
fraction of missing data in the final maps.

\subsubsection{Astrometry Check}
\label{sec:astrometry}
As discussed in detail in \citet{schaffer11}, the reconstruction of the pointing (the instantaneous
sky location viewed by every detector at every time sample) for the SPT is based on daily measurements
of the Galactic HII regions RCW38 and Mat5a, supplemented with information from 
thermal, linear displacement, and tilt sensors in the telescope. The typical precision in this 
reconstruction is 7 arcsec (as measured by the rms variation in bright source positions over
many individual observations of a field). The overall astrometric solution for SPT maps is 
refined by comparing to source positions in the Australia Telescope 20 GHz Survey
(AT20G) catalog \citep{murphy10}, which are tied to very long-baseline interferometry calibrators
and are accurate at the 1-arcsec level. When we apply this technique to the LMC and SMC
fields, we find small (10-15 arcsec) but statistically significant offsets between the original SPT 
positions and the AT20G positions. We correct these offsets by simply redefining the map centers.
The final map centers (which we use to reproject \planck\ data onto the SPT grid and which
we publish in the final combined map FITS files) are R.A.~$79.9906^\circ$, decl.~$-68.4984^\circ$
for the LMC and R.A.~$14.9849^\circ$, decl.~$-72.4994^\circ$ for the SMC. Based on the analysis in 
\citet{schaffer11} we expect that, after this correction, the astrometry is good to roughly 2~arcsec
rms.

\subsection{\planck}
\label{sec:planck}
The primary science goal of the \planck\ satellite \citep{planck13-1}, launched in 2009 by the 
European Space Agency, was to map the CMB over the full sky in nine bands, ranging in 
wavelength from 350~$\mu$m to 1~cm. 
In this work, we use publicly available \planck\ data in the three wavelength
bands that closely overlap with the three SPT bands. These are three longest-wavelength or 
lowest-frequency bands on the \planck\ High-Frequency Instrument (HFI) and have nominal center 
wavelengths of 1.4, 2.1, and 3.0~mm (nominal center frequencies of 217, 143, and 100~GHz).
The instrument beam or point-spread function in these three bands is close to Gaussian
and azimuthally symmetric, with 
FWHM equal to 5.0, 7.1, and 10.0 arcmin at 1.4, 2.1, and 3.0~mm, respectively. 

The \planck\ HFI time-ordered data are combined into maps using an approximation to the 
minimum-variance solution \citep{planck11-6}, in contrast to the technique of filtering and 
naive bin-and-averaging used to make the SPT maps. This results in maps that are unbiased
estimates of the true sky signal at all scales except for the effect of the instrument beam
and pixelization, 
and the DC component of the maps, which is set to zero in the HFI mapmaking
procedure (\citealt{planck13-8}; for more discussion of the zero-point treatment, see
Section~\ref{sec:galforeg}).
Thus, the angular response functions appropriate for the \planck\ HFI
maps are simply the convolution (or Fourier-space product) of the
instrument beams and the known pixel window function. For more details
on the \planck\ HFI instrument and data, see \citet{lamarre10}, \citet{planck13-6}, and \citet{planck15-8}.

To create \planck\ HFI maps of the Magellanic Clouds that match the SPT maps described in the
previous section, we first take the publicly available full-mission maps\footnote{Downloaded from 
the NASA/IPAC Infrared Science Archive: \url{http://irsa.ipac.caltech.edu/data/Planck/release\_2/all-sky-maps}.}
in each of the three bands and resample them from their native pixelization onto the
0.25-arcmin oblique Lambert equal-area azimuthal (ZEA) projection used for the SPT maps.
For the R.A./decl. center of the target projection, we use the center of the SPT maps derived 
from the astrometry cross-check with the AT20G survey (see Section~\ref{sec:astrometry} for details).
The HFI maps are stored using the full-sky
HEALPix\footnote{\url{http://healpix.sourceforge.net}} pixelization scheme, with the HEALPix $N_\mathrm{side}$
parameter set to 2048, leading to $12 \times 2048^2$ pixels over the full sky, or a pixel scale
of 1.7~arcmin. In the resampling to the 0.25-arcmin flat-sky grid, we oversample each 0.25-arcmin
pixel by a factor of four to reduce the effect of resampling artifacts. 

The \planck\ maps in the ZEA projection are then matched to the resolution of the SPT
maps by dividing the \planck\ maps in 2d Fourier space by the ratio of the \planck\ beam
to the SPT beam in the closest observing band. 
The \planck\ beams used in this operation are the 
product of the publicly available measured instrument beams and the HEALPix 
$N_\mathrm{side}=2048$ pixel window function.
This Fourier-space operation is equivalent to 
deconvolving the \planck\ beam from the map and convolving the result with the corresponding
SPT beam. At small enough scales (high enough wavenumber $k$), this ratio becomes small 
enough to cause numerical issues---and becomes increasingly uncertain as the fractional 
\planck\ beam uncertainties grow larger---so we artificially roll off the ratio at low values
of the \planck\ beam ($B(k) < 0.005$). This roll-off is taken into account when we combine
the SPT and \planck\ maps (see Section \ref{sec:weights} for details).

We also create null \planck\ HFI maps---using the publicly available \planck\ half-mission 
maps---to combine with the null SPT maps described 
in Section \ref{sec:sptdata}. We make the null \planck\ maps by subtracting one half-mission
map from the other half-mission map (divided by two) in each band, then resampling to the
ZEA projection, and deconvolving the \planck-SPT beam ratio,
as done for the signal maps. 
As was the case in the SPT
null maps, there are small artifacts in the \planck\ null maps at the location of the brightest
regions of the two fields, and, as with the SPT null maps, the artifacts are at the percent 
level or below.

To combine these \planck\ HFI maps with the SPT maps described in Section \ref{sec:spt}, we 
need an estimate of the noise properties of the SPT-beam-matched \planck\ maps. 
The noise in \planck\ HFI maps is uncorrelated
between pixels (white) to a very good approximation \citep{planck11-6}, so the Fourier-domain
\planck\ map noise in a uniform-coverage region is well approximated by a single value at all 
$k$ values or angular scales. Thus, the Fourier-domain map noise in one of the SPT-beam-matched
\planck\ maps is this value divided by the ratio of the \planck\ and SPT beams, under the
assumption that the \planck\ noise is uniform over the map.

The \planck\ coverage in the SMC field is quite uniform, only
varying by $\pm 12 \%$ across the field (corresponding to $\pm 6\%$ variations in noise).
The LMC field is near the south ecliptic pole, and the \planck\ observing strategy results in 
regions of very high coverage near the ecliptic poles. Approximately 25\% of pixels in the LMC
field are in such a region (defined as 50\% higher coverage than the mode of the distribution in
the rest of the field). Using a single value for the noise across the field will result in a slightly 
suboptimal combination of SPT and \planck\ data for the high-weight regions. No bias results
from this approximation, and the variation of \planck\ noise across the field will be properly represented
in the combined SPT+\planck\ null maps. For both fields, we estimate the \planck\ noise by 
taking the square root of the mean of the variance values for all pixels in the region covered 
by SPT. The pixel variance values are provided by the \planck\ team in the same files as the maps.

\section{Combining Data From SPT and {\sl PLANCK}}
\label{sec:combine}
There are two main steps in the process of optimally combining the SPT and \planck\ maps
described in previous sections. First, the maps are relatively calibrated (or, more specifically, the 
SPT map is adjusted to match the \planck\ maps) and converted from CMB fluctuation temperature
to brightness or specific intensity (in units of MJy sr$^{-1}$) at a fiducial observing wavelength
and for an assumed source spectrum.
Then the maps from the two instruments are combined into a single map using inverse-variance
weights calculated from the noise estimate for each instrument
in each field and band. Each of these steps is described in greater detail below.

\subsection{Absolute Calibration}
\label{sec:cal}
Before the SPT and \planck\ HFI data can be meaningfully combined into a single map, care 
must be taken to ensure that the two data sets are consistently calibrated---that is, that a true
sky signal would produce the same amplitude of response in both data sets (up to differences 
in the angular response function of the two instruments). Maps from both instruments are stored
in units of CMB fluctuation temperature, i.e., the variation in temperature of a blackbody with
mean temperature 2.73K that would produce the detected signal. The absolute calibration of the
\planck\ maps used in this work is taken from the annual modulation of the CMB dipole due to the 
motion of the satellite around the solar system barycenter \citep{planck15-8}. The fractional 
statistical uncertainty on this calibration is significantly less than 1\%. This 
calibration can be checked by comparing the CMB power spectrum measured with \planck\ to
the CMB power spectrum measured by the \wmap\ team, who also calibrate their data off of 
the modulation of the CMB dipole, using internal \wmap\ measurements. The CMB power 
spectrum measurements from the two instruments agree to better than 1\% in power
(0.5\% in CMB fluctuation temperature, \citealt{planck15-1}).

The SPT absolute calibration is obtained by matching the small-scale 
(high-multipole) CMB power spectrum measured with SPT and published in \citet{george15}
with the \planck\ CMB power spectrum over the same multipole range ($670 < \ell < 1170$,
where $\ell$ is multipole number and, over small patches of sky, is equivalent to angular
wavenumber $k$ as defined in Section \ref{sec:sptxfer}). The fractional statistical uncertainty on 
this calibration is roughly 2.5\% (in temperature) at 1.4~mm and 1.0\% (in temperature) at
2.0 and 3.2~mm.

\begin{figure}
\includegraphics[width=3.5in]{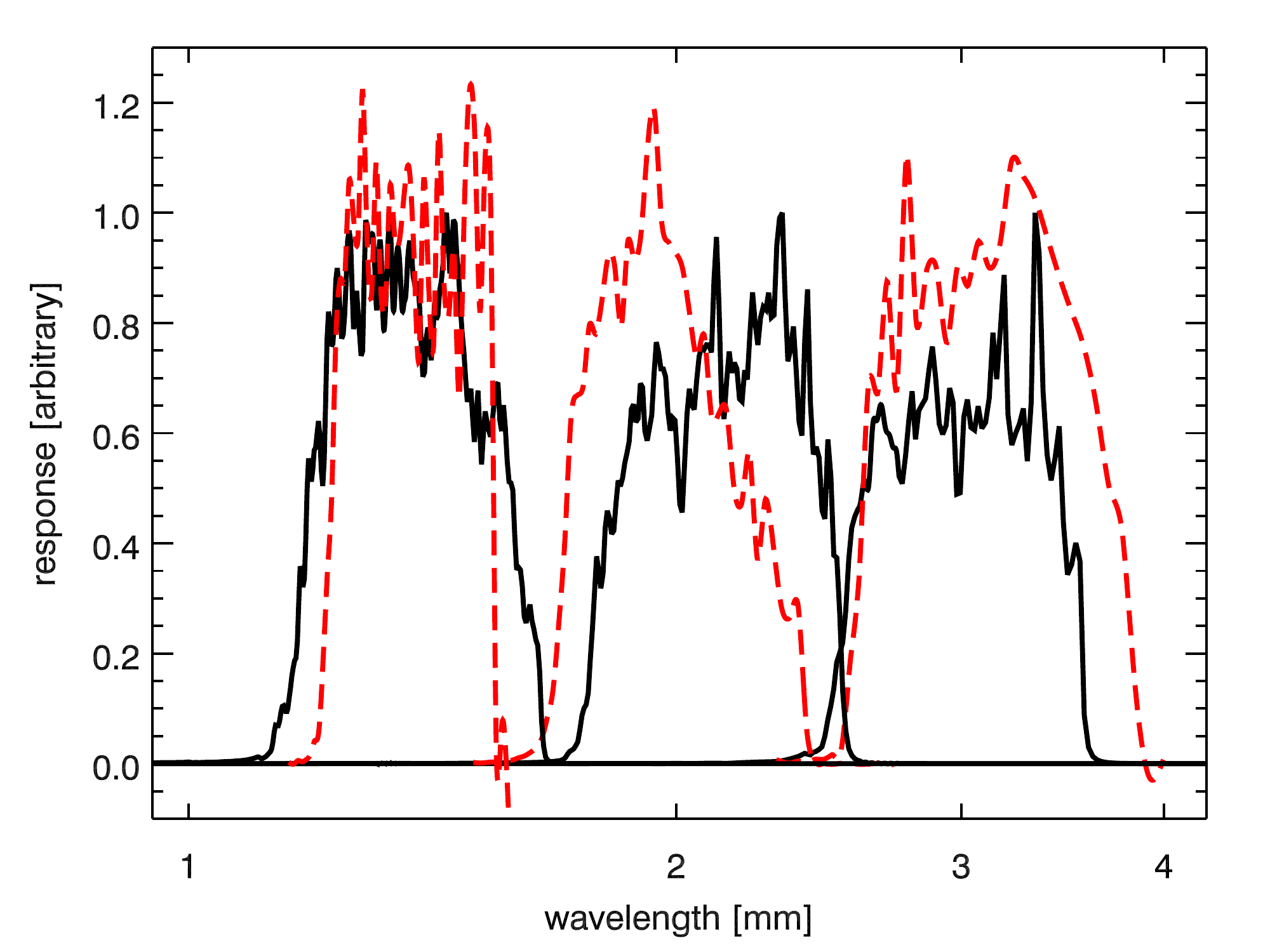}
\caption{Bandpass functions, or instrument response as a function of wavelength, for 
SPT-SZ and the lowest three bands of \planck\ HFI. \planck\ bands are shown by the
solid black lines, while SPT-SZ bands are shown by the dashed red lines.
The normalization of the bandpasses is arbitrary.
}
\label{fig:bands}
\end{figure}

\subsection{Spectral Matching}
\label{sec:conv}
As discussed in the previous section, the absolute calibration of both the SPT and \planck\ 
data used here is based on a source with an emission spectrum described by
fluctuations around a 2.73K blackbody, i.e., 
$I(\lambda) \propto dB/dT(\lambda,2.73K)$, where $B(\lambda,T)$ is the Planck blackbody function. 
If the SPT and \planck\ bands were infinitely narrow, or if they had finite width but were 
identical in response as a function of wavelength (or bandpass), the calibration step described above would
be sufficient for matching the SPT and \planck\ maps of a source with an arbitrary emission
spectrum. In reality, the SPT and \planck\ bands have fractional widths of order 30\%, and there are
small but significant differences in the bandpass functions for the two instruments, 
as shown in Figure \ref{fig:bands}. (The publicly available \planck\ bands were
downloaded from the same server as the maps and instrument beams.)

Because of this small bandpass mismatch, and because the emission from the 
Magellanic Clouds is not expected to have a $dB/dT(\lambda,2.73K)$ spectrum, 
we need to apply some further correction factor to match
the SPT and \planck\ responses to the emission from the Magellanic Clouds.
The size of that correction depends on the spectral energy distributions (SEDs) 
at each point within
 the LMC and SMC and the different SPT and \planck\ bandpass functions.
 To choose the appropriate
spectral matching factor, we need some prior information on the SEDs of the Magellanic Clouds.
Fortunately, the SEDs can be approximated by power laws with a limited range of index. 
In the following section, we use \planck\ data in the bands under investigation here and
in the neighboring HFI bands to estimate the SEDs of the LMC and SMC at the angular
scales accessible to \planck.

\begin{figure*}
\begin{centering}
\subfigure[LMC]{\label{fig:alpha_lmc} 
 \includegraphics[width=3.25in]{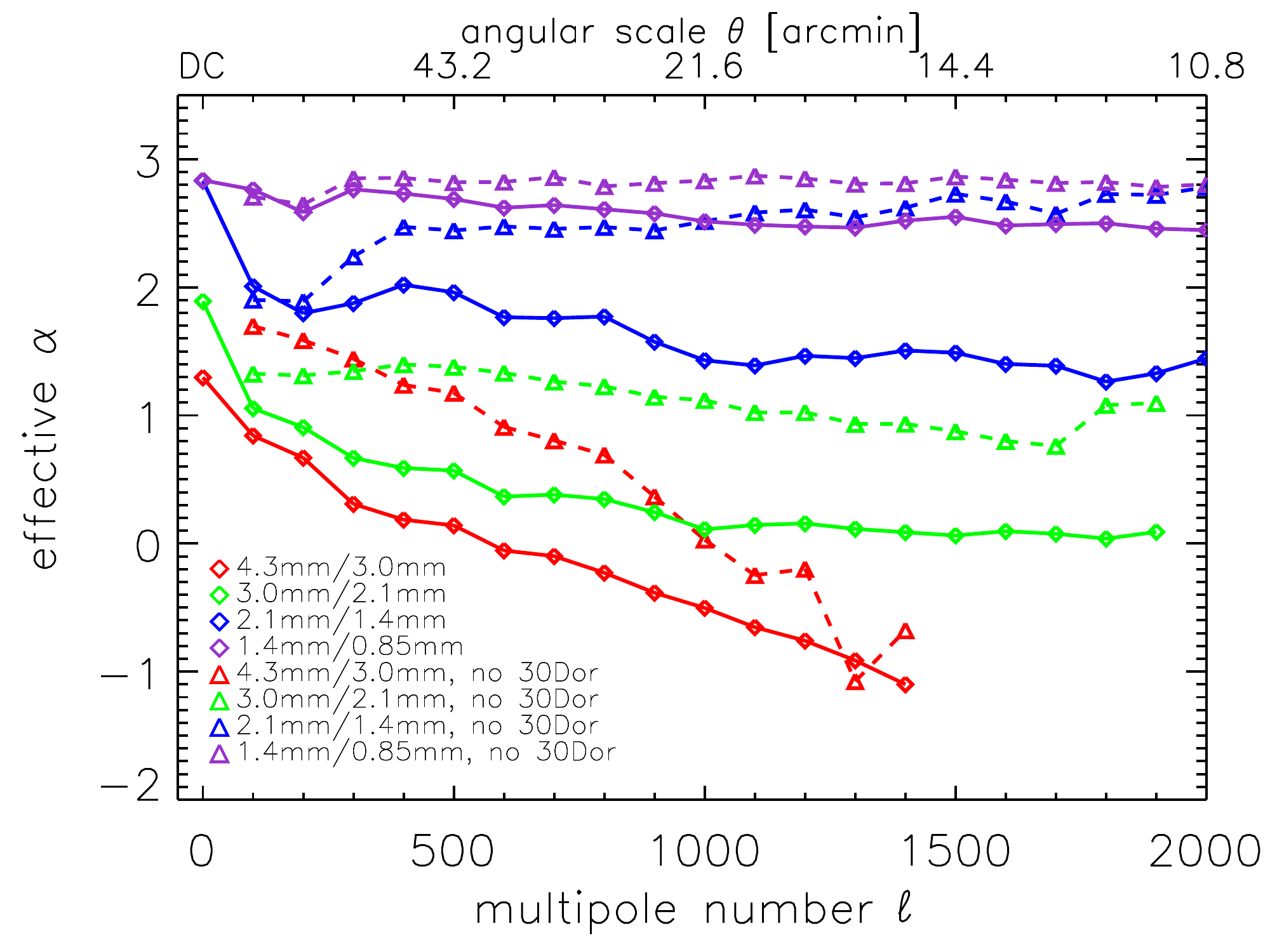}}
\subfigure[SMC]{\label{fig:alpha_smc} 
 \includegraphics[width=3.25in]{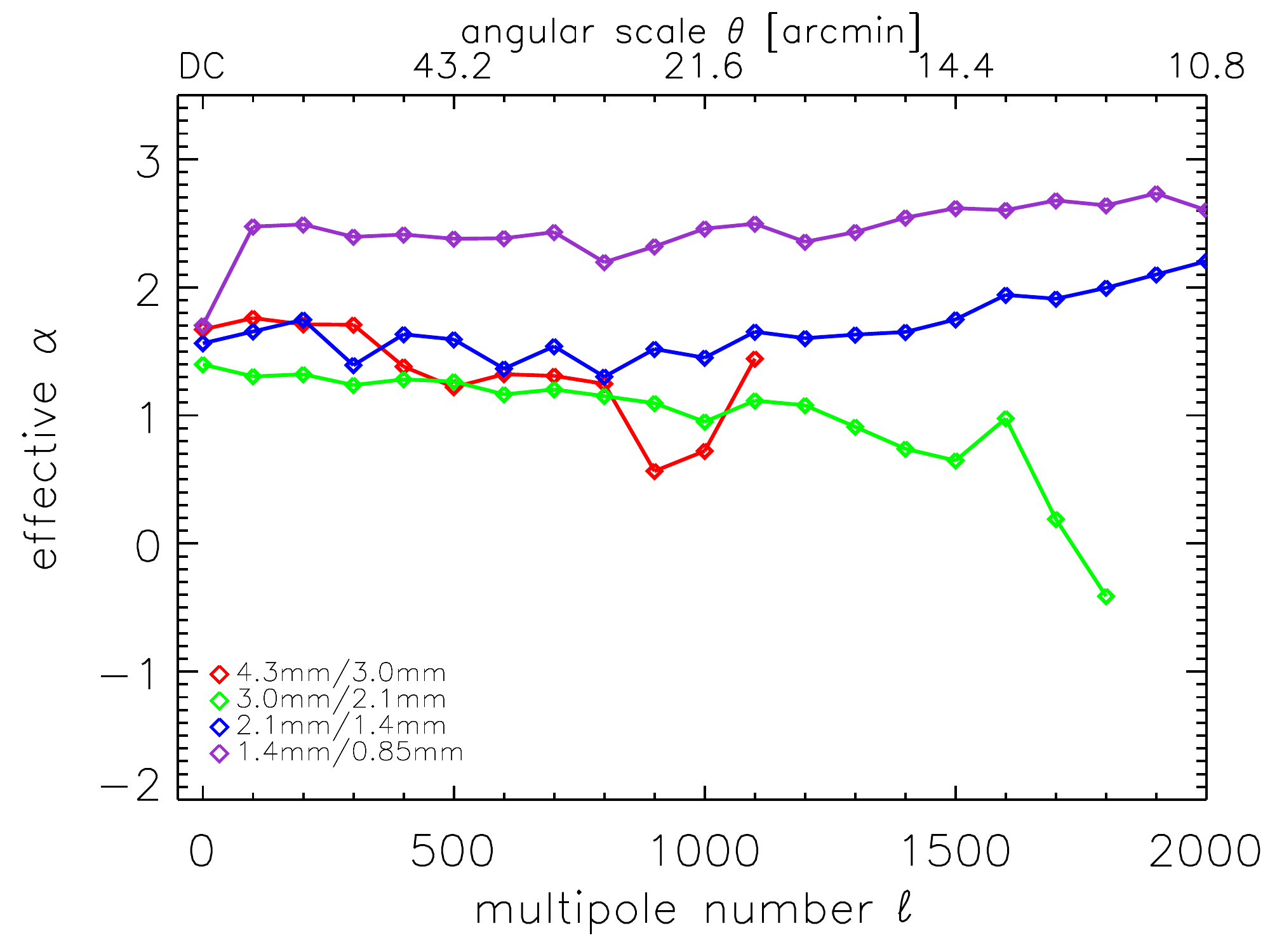}}
\caption{Effective power-law index $\alpha$ as a function of angular scale or multipole number $\ell$ for
four combinations of \planck\ bands in the LMC ({\it Left Panel}) and 
SMC ({\it Right Panel}). The $\ell=0$ value is calculated using the ratio of 
integrated emission from the full LMC or SMC region, while all other 
values are calculated using the ratio of the (beam-corrected) power spectra of the \planck\ 
maps in each pair of bands. In the left panel, the 
solid lines show the effective $\alpha$ from the entire LMC region, while the dashed
lines show the effective $\alpha$ with 30~Doradus excluded. (In all cases, lines are only drawn for 
values of $\ell$ at which the beam window function for the longest-wavelength band
is above 5\% of peak value and for which effective $\alpha$ is well-defined.)}
\end{centering}
\label{fig:alpha}
\end{figure*}

\subsubsection{Spectral Energy Distributions of the Magellanic Clouds from \planck-only Data.}
\label{sec:planckonly}
In this section, we use \planck\ data from 0.85~mm to 4.3~mm to estimate the SEDs
of the LMC and SMC at \planck\ angular scales. 
One complication to this process 
is evident from \planck-only maps of the Magellanic Clouds in Figure~1 of
\citet{planck11-17}. For example, in the LMC, the dust emission (traced by the 0.35~mm 
or 857~GHz map) is quite diffuse and covers the entire region, while the synchrotron
emission (traced by the 10.5~mm or 28.5~GHz map) is concentrated in bright regions
such as 30 Doradus.
This makes it unlikely 
that a single SED will be sufficient to describe the emission across the full extent 
of the LMC and SMC.

A more quantitative view of this issue is provided in Figure~\ref{fig:alpha}, 
which shows the effective power-law index $\alpha$ (defined assuming specific intensity 
$I \propto \lambda^{-\alpha}$) in the LMC and SMC
as a function of angular scale or multipole number $\ell$ for
four combinations of bands: 4.3~mm$/$3.0~mm, 3.0~mm$/$2.1~mm, 2.1~mm$/$1.4~mm,
and 1.4~mm$/$0.85~mm. The DC ($\ell=0$) value is calculated using the ratio of 
integrated emission for each pair of bands given in Table~2 of \citet{planck11-17}, while all other 
values are calculated using the ratio of the power spectra of the \planck\ maps in each pair of bands
(corrected by the square of the ratio of beam window functions of the two bands). The 
solid lines show the effective $\alpha$ from the entire LMC and SMC regions, while the dashed
lines in the left panel show the effective $\alpha$ for the LMC with 30~Doradus excluded. (Note that values for a particular
pair of bands are only shown for values of $\ell$ at which the beam window function of both
bands is greater than 5\% of the peak value 
and for which effective $\alpha$ is well-defined.) It is clear that the shortest 
wavelengths are dominated by dust emission at all scales and in all regions, while free-free
and synchrotron emission contribute significantly to the small-scale emission at longer
wavelengths, particularly in the LMC when 30 Doradus is included. 

\subsubsection{Matched SPT-\planck\ Maps for Different SEDs}
In light of the observed variation in $\alpha$, we choose to make several different combinations
of \planck\ and SPT maps, each appropriate for a different assumed source spectrum.
For each such combination, we convert all six maps (three \planck\ maps, three SPT maps)
from units of CMB fluctuation temperature to units of MJy~sr$^{-1}$ assuming a power-law emission spectrum 
$I(\lambda) \propto \lambda^{-\alpha}$, evaluated at the nominal central wavelength
of the \planck\ bands. That is, all six maps are scaled such that they represent
an estimate of the specific intensity of a $\lambda^{-\alpha}$ source at a wavelength of 
1.4, 2.1, or 3.0~mm. 
The conversion
from CMB fluctuation temperature to specific intensity at a given reference wavelength
is performed using the formalism described in Section 3.2.3 of \citet{planck13-9} and the bands shown in
Figure \ref{fig:bands}. The null maps and the Fourier-space noise estimates for each
instrument are converted using the same factors used for the signal maps.

Guided by the range of effective spectral indices for the LMC seen in Figure~\ref{fig:alpha}, 
we create five different versions of the SPT and \planck\ maps, each appropriate for combining
the two sets of maps if the true sky emission has a particular spectral index. The values of spectral
index we assume for the five versions are $\alpha = \{-1.0,0.0,1.0,2.0,3.0\}$. 
The magnitude of the error incurred by using a map created assuming an incorrect value of $\alpha$ depends on
the width of the bands and the fiducial wavelength used for the conversion. The fractional widths of
the three \planck\ bands are similar, and the fiducial wavelengths
we use are the nominal \planck\ band centers. For these reasons, the difference in conversion
factors in the \planck\ bands is small ($<10\%$) for the range of spectral indices used. This is also
the case for the SPT 1.4~mm and 3.2~mm bands, which are similar to the corresponding \planck\
bands in placement and width. However, the SPT 2.0~mm band is significantly offset from the
\planck\ 2.1~mm band, and the factors to convert data in that band from CMB fluctuation temperature to MJy~sr$^{-1}$ at the
nominal \planck\ band center vary by 30\% between $\alpha=-1$ and $\alpha=3.$ This is the
strongest motivation for creating multiple sets of combined maps. If a user of the final data 
products needs a map appropriate for a non-integer spectral index (or an index outside the range
we use) and desires conversion accuracy better than 5\% (roughly the difference between the 
SPT 2.0~mm conversion factors assuming $\alpha=\alpha_0$ and assuming $\alpha=\alpha_0 \pm 1$), 
we suggest interpolating between maps or extrapolating.

\subsubsection{Beam-filling vs.~Point-like Sources}
A further complication to the calculation of conversion factors between CMB temperature 
and MJy~sr$^{-1}$ is the question of beam-filling vs.~point-like sources. For diffraction-limited 
optical systems that couple to a single mode of radiation per polarization, the product of 
telescope area and beam solid angle ($A\Omega$ or \'etendue) is equal to $\lambda^2$. Both SPT
and \planck\ operate in or near this single-moded, diffraction limit for the bands considered
here \citep{padin08,planck-prelaunch-1}. In the limit of constant telescope aperture illumination as 
a function of wavelength, the $A \Omega$ from a point-like source will be constant, while the 
$A \Omega$ for a beam-filling source will go as $\lambda^2$. The power received by an optical
system from a source scales directly with $A \Omega$, so in this limit the conversion factor
calculated for a point source with spectral index $\alpha$ will be equivalent to the conversion
factor for a beam-filling source with spectral index $\alpha + 2$. For the reasons discussed above,
this distinction matters significantly only for the SPT 2.0~mm band, where it is up to a 15\% 
difference. All the conversion factors we use here assume beam-filling sources, mainly because
that is how the bands and absolute calibration were measured for both instruments 
\citep{schaffer11,planck13-9}. There are few isolated features at sub-arcminute scales 
in the Magellanic Clouds (e.g., \citealt{meixner13}, Figure 14), so this is a safe choice for
the SPT bands; the assumption is less safe for \planck, but the differences between beam-filling 
and point-source conversion factors for \planck\ are much smaller (at the percent level).

\subsection{Combining Maps Using Inverse-variance Weighting}
\label{sec:weights}
Once the SPT and \planck\ maps are in a common set of units and are consistently
calibrated, and we have estimates of the noise for 
both maps, it is straightforward to combine them in an optimal (minimum-variance)
way using inverse-variance weights. 
The noise estimates are not perfect, and the combined maps are thus not
truly optimal. The degree to which the combined maps are sub-optimal
depends on the fidelity of the noise estimates and the
assumptions underlying these estimates, particularly that of uniform noise across 
the entire map.
As discussed in previous sections, the SPT
noise properties are very uniform across the map in both the LMC and SMC fields, 
while the \planck\ noise properties are very uniform in the SMC map but not in 
the LMC map. (A fraction
of the pixels in the LMC have \planck\ noise that is significantly lower than the mean.)
The SPT-\planck\ combination will be slightly sub-optimal for these map regions, but
there is no signal bias; furthermore, the combined SPT-\planck\ null maps properly reflect
this variation in noise in the LMC field. 

To combine the maps in a given band, we first Fourier transform each map.
At each point in 2d Fourier space $\bk = [k_x,k_y]$, we define a weight for 
each map 
\beq
W(\bk) = N^{-2}(\bk), 
\eeq
where $N(\bk)$ is the Fourier-space noise estimate.
Recall that for \planck\, we assume white noise in the raw maps, so that the Fourier-space
noise in the SPT-beam-matched maps will be proportional to the ratio of the beams:
\beq
N_\mathrm{Planck}(\bk) \propto B_\mathrm{SPT}(\bk) / B_\mathrm{Planck}(\bk).
\eeq
This ratio is a monotonically increasing function of $k$ and reaches a value
of 10 at approximately $k=3700$ at 1.4~mm, $k=2500$ at 2.1~mm, and 
$k=1900$ at 3.0~mm. Hence, the \planck\ weights will be down by a factor of
100 at these $k$ values compared to $k=0$.
Meanwhile, the SPT maps have had the filter response function deconvolved, so the 
noise in these maps is given by
\beq
N_\mathrm{SPT}(\bk) = N_\mathrm{SPT, orig}(\bk) / F_\mathrm{SPT}(\bk),
\eeq
where $N_\mathrm{SPT, orig}(\bk)$ is the estimate of the noise from the original maps, 
and $F_\mathrm{SPT}(\bk)$ is the SPT filter response function. From this it is clear that
the \planck\ weights will be proportional to the square of the ratio of the \planck\ beam to the
SPT beam, while the SPT weights will be reasonably flat (depending on the noise properties
in the original maps) except for the regions of Fourier space strongly affected by the filtering,
which will have much lower weight. 
We manually zero the \planck\ weights 
at any values of \bk\ at which the value of the \planck\ beam is lower than 0.005,
and we manually zero the SPT weights at values of \bk\ at which the azimuthally symmetric part of the SPT filter
response is less than 0.01.

\begin{figure}
\includegraphics[width=3.5in]{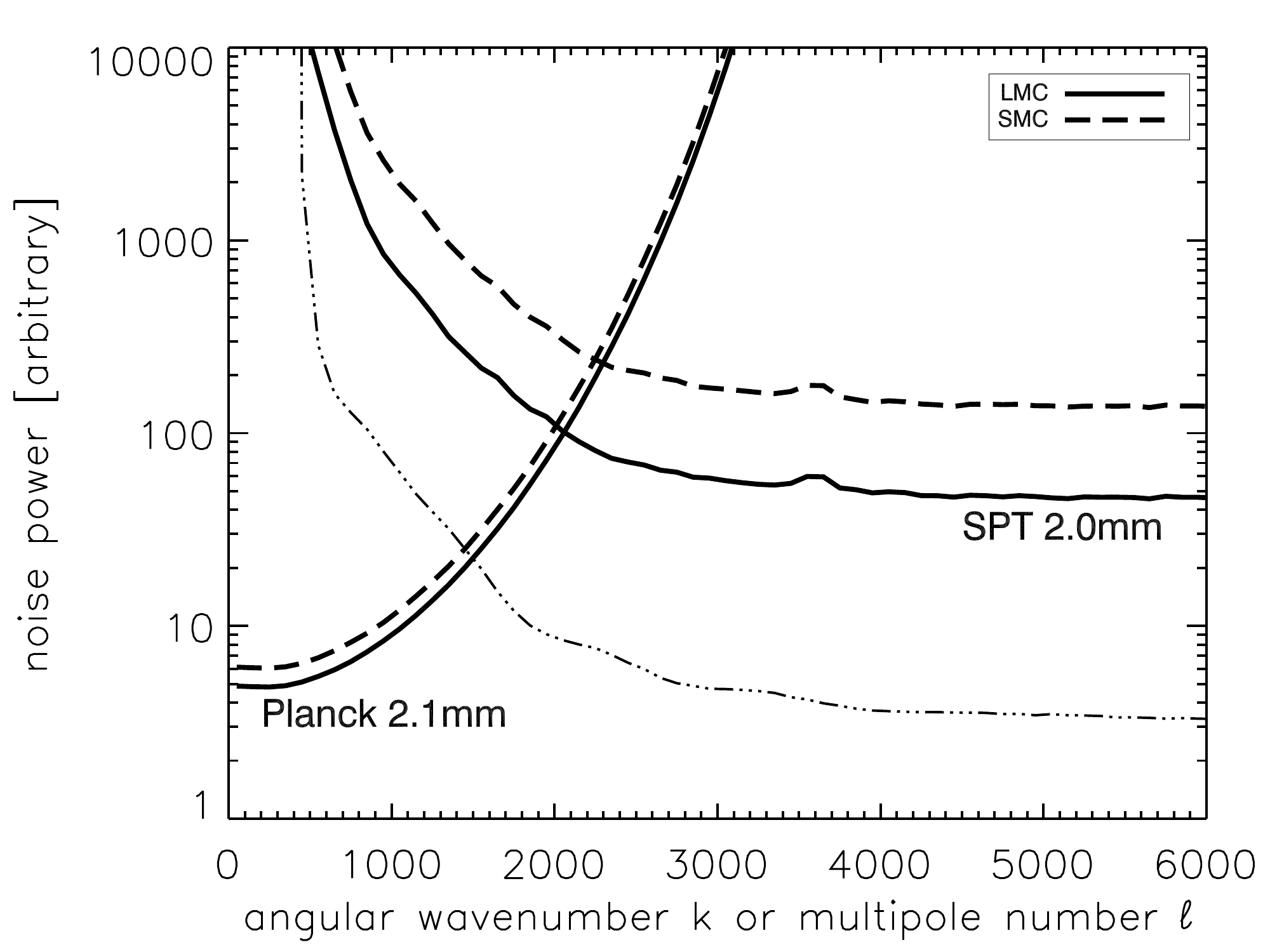}
\caption{Noise power as a function of angular wavenumber $k$ or multipole number $\ell$ for the 
\planck\ 2.1~mm band and the SPT 2.0~mm band. The solid lines show the noise power in the LMC
field; dashed lines show the noise power in the SMC field. The SPT noise is measured after deconvolving
the filter response function, while the \planck\ noise is measured after matching to the SPT beam 
(see Section~\ref{sec:weights} for details).
In all cases, the assumed spectral index for
the emission is $\alpha=2.0$. The small bump in the SPT noise at $k=3600$ is due to a noise
feature associated with the 6-arcmin elevation step between scans (see Section~\ref{sec:sptobs} for details).
The thin dot-dashed line shows the 2.0~mm noise power for a typical field in the SPT-SZ survey 
(calculated in the same manner as for the LMC and the SMC curves in this plot).
}
\label{fig:noisevscale}
\end{figure}

The azimuthally averaged values of $N^2_\mathrm{Planck} (\propto W^{-1}_\mathrm{Planck})$ 
and $N^2_\mathrm{SPT} (\propto W^{-1}_\mathrm{SPT})$ as a function
of $k$ in both fields are shown for the 2.1/2.0~mm bands and an assumed spectral index $\alpha=2.0$ in 
Figure~\ref{fig:noisevscale}. For comparison, we also show $N^2_\mathrm{SPT}$ at 2.0~mm from a typical
field in the 2500-square-degree SPT-SZ survey (in which the typical field was observed a factor
of roughly 10 longer than the total LMC observations used in this work).
Plots of $N^2_\mathrm{Planck}$ and $N^2_\mathrm{SPT}$ 
in the 1.4~mm and 3.0~mm bands look similar, except 
that the \planck/SPT crossover point is at higher $k$/$\ell$ for the 1.4~mm band.

The combined map in Fourier space is then simply
\beq
M_\mathrm{comb}(\bk) = \frac{W_\mathrm{SPT}(\bk) M_\mathrm{SPT}(\bk) + 
W_\mathrm{Planck}(\bk) M_\mathrm{Planck}(\bk)}{W_\mathrm{SPT}(\bk) + W_\mathrm{Planck}(\bk)},
\label{eqn:comb}
\eeq
where $M(\bk)$ indicates a Fourier-space map. We note that there are no modes 
for which both the \planck\ and the SPT weights have been manually zeroed; for all modes
of interest, Equation \ref{eqn:comb} is well defined.
We then inverse Fourier transform to 
obtain the combined map in real space $M_\mathrm{comb}(x,y)$.

Finally, to limit the effect of noise near the beam scale, we make three smoothed versions
of each combined map at 1.4 and 2.1~mm and two smoothed versions of each map at 
3.0~mm. For the 1.4 and 2.1~mm maps, we make a version that is only slightly smoothed
compared to the SPT beam: we convolve these maps with the difference between the 
SPT beam at that wavelength and a 1.5-arcmin FWHM Gaussian. (We do not make this 
version of the 3.0~mm map, because the SPT beam itself is roughly a 1.7~arcmin FWHM 
Gaussian.) For maps in all wavelength bands, we make versions with 2.0 and 2.5-arcmin
resolution by convolving the combined SPT-resolution map with the difference between
the SPT beam and a 2.0 or 2.5-arcmin FWHM Gaussian. We provide maps at different resolutions
in an attempt to balance the requirements of low noise and high angular resolution. End users
of these maps that are interested in the smallest-scale features of the Magellanic Clouds
should use the highest-resolution maps and pay the penalty of slightly higher noise, while
users interested in, for example, performing few-arcminute scale photometry on the maps
should use the lower-resolution versions with reduced noise. Users can also of course 
perform their own filtering on the data---for example, loading one of the 1.5-arcmin-resolution
maps into the SAOImage ds9 software\footnote{\url{http://ds9.si.edu}}
and applying the Gaussian smoothing kernel with 
a kernel radius of seven pixels very closely reproduces the 2.5-arcmin-resolution version
of that map.

\subsection{Treatment of Galactic Foregrounds and Map Zero Level}
\label{sec:galforeg}
A primary use of the combined SPT-\planck\ maps described in this work is
expected to be aperture photometry on localized regions of the LMC and SMC. In this
application, any emission from sources other than the Magellanic Clouds will either
act as a source of bias or extra variance, depending on the whether that emission varies 
significantly over the LMC or SMC field. In particular, the mean value of any source of 
non-Magellanic-Clouds emission will act as a bias, so we attempt to remove the mean of 
all such sources from the combined SPT-\planck\ maps.

In our combining procedure, the DC ($k=0$) weight for the SPT maps is always
set to zero, so the mean across the combined maps will be equal to the mean of the \planck\
map. As discussed in Section~\ref{sec:planck}, the \planck\ maps are constructed so that the 
mean across the full sky is identically zero. However, the expected mean of the Galactic signal
and the cosmic infrared background (CIB) are manually added back to the \planck\ maps before
public release---though the mean emission from the 2.73K CMB is not \citep{planck13-8}. 
Though the CIB monopole
is much smaller than the mean Galactic signal, we subtract it from the maps nonetheless. The 
variance of the residual CIB fluctuations is completely negligible compared to noise and CMB
fluctuations, and we ignore it.

The signal from our own galaxy is not a random field like the CMB
or CIB, so we do not subtract the full-sky mean from our maps but rather an estimate of the mean
in the direction of the LMC or SMC. We estimate this mean in each \planck\ band by taking the 
mean signal quoted at 0.35~mm (857~GHz) in \citet{planck11-17} toward each region and scaling
by the Galactic cirrus spectral energy distribution quoted in \citet{planck12-12}.
This assumes the Galactic foreground signal is dust-dominated across all the frequencies 
treated here, an assumption supported by Figure~4 of \citet{planck11-17}.
The residual variance across the LMC and SMC is estimated in \citet{planck11-17}
to be $\ll 10^{-3} \ \mathrm{MJy} \ \mathrm{sr}^{-1}$ rms in all bands and in both regions, 
a factor of at least 10 below the CMB fluctuation level (see Section~\ref{sec:cmbnoise}), and
we ignore this source of variance as well. The variance from CMB fluctuations is discussed in
detail in Section~\ref{sec:cmbnoise}.

\begin{figure*}
\begin{centering}
\subfigure[\planck-only, SPT-only, and \planck+SPT maps of the 7-by-7-degree LMC field]{\includegraphics[width=2.5in]{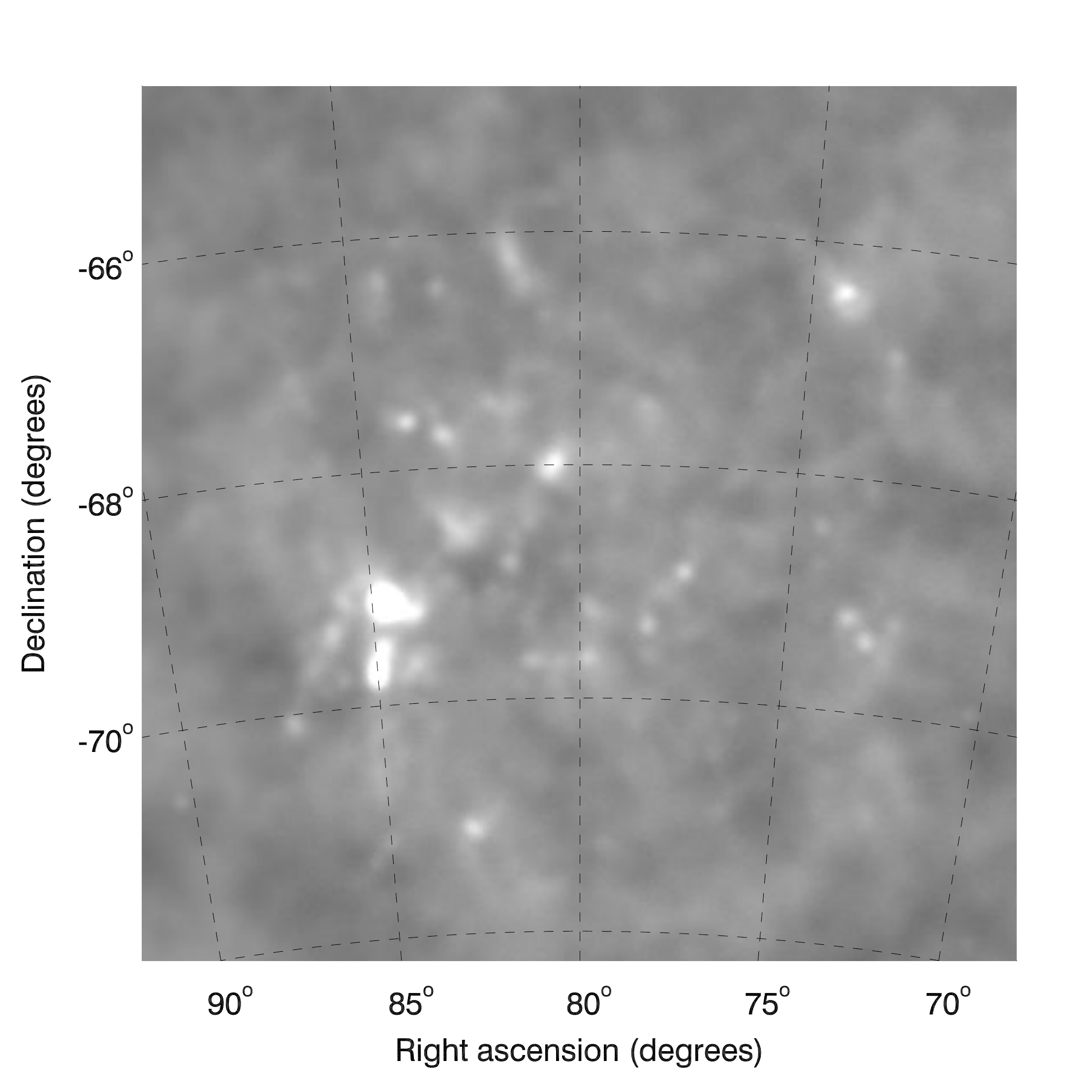}
\includegraphics[width=2.5in]{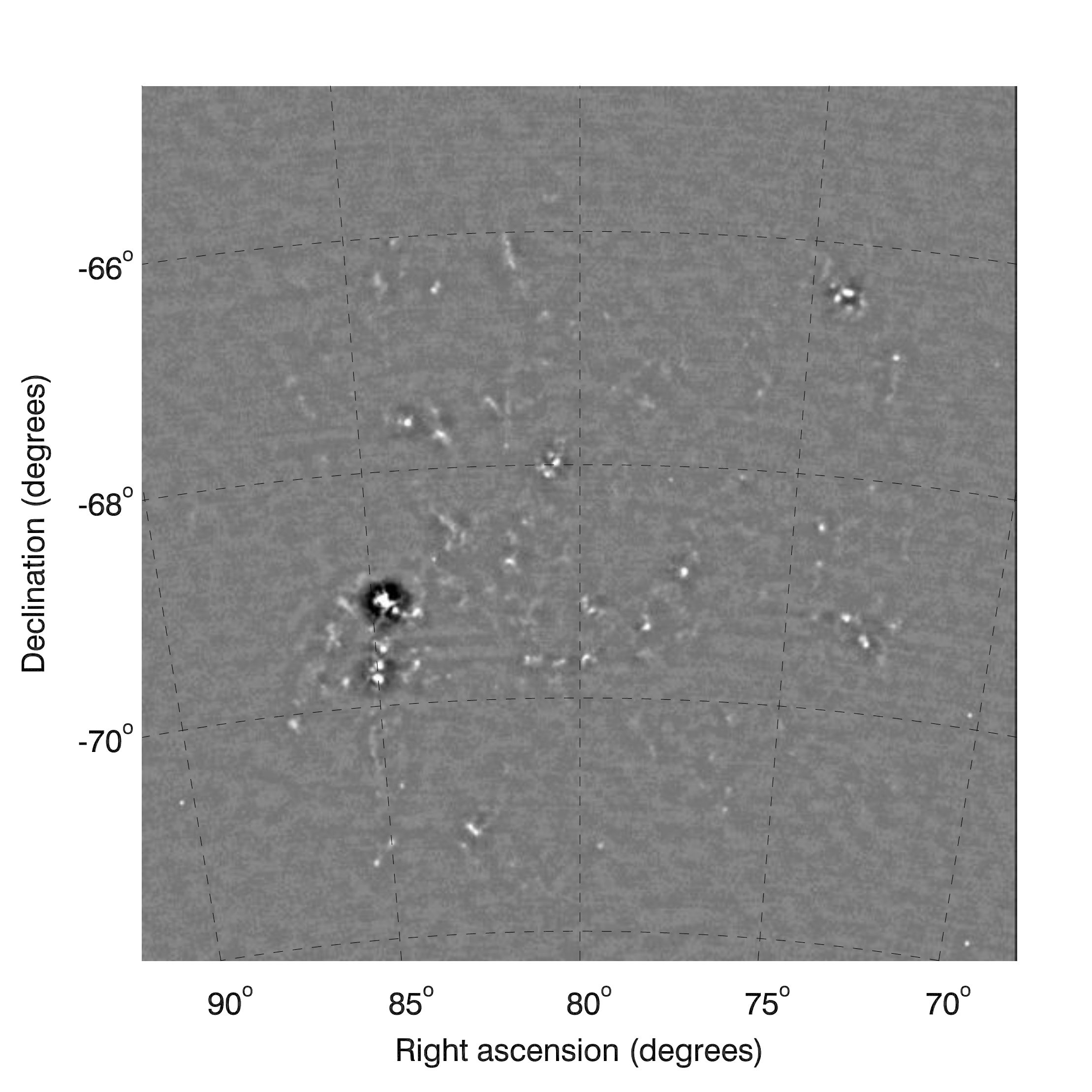}
\includegraphics[width=2.5in]{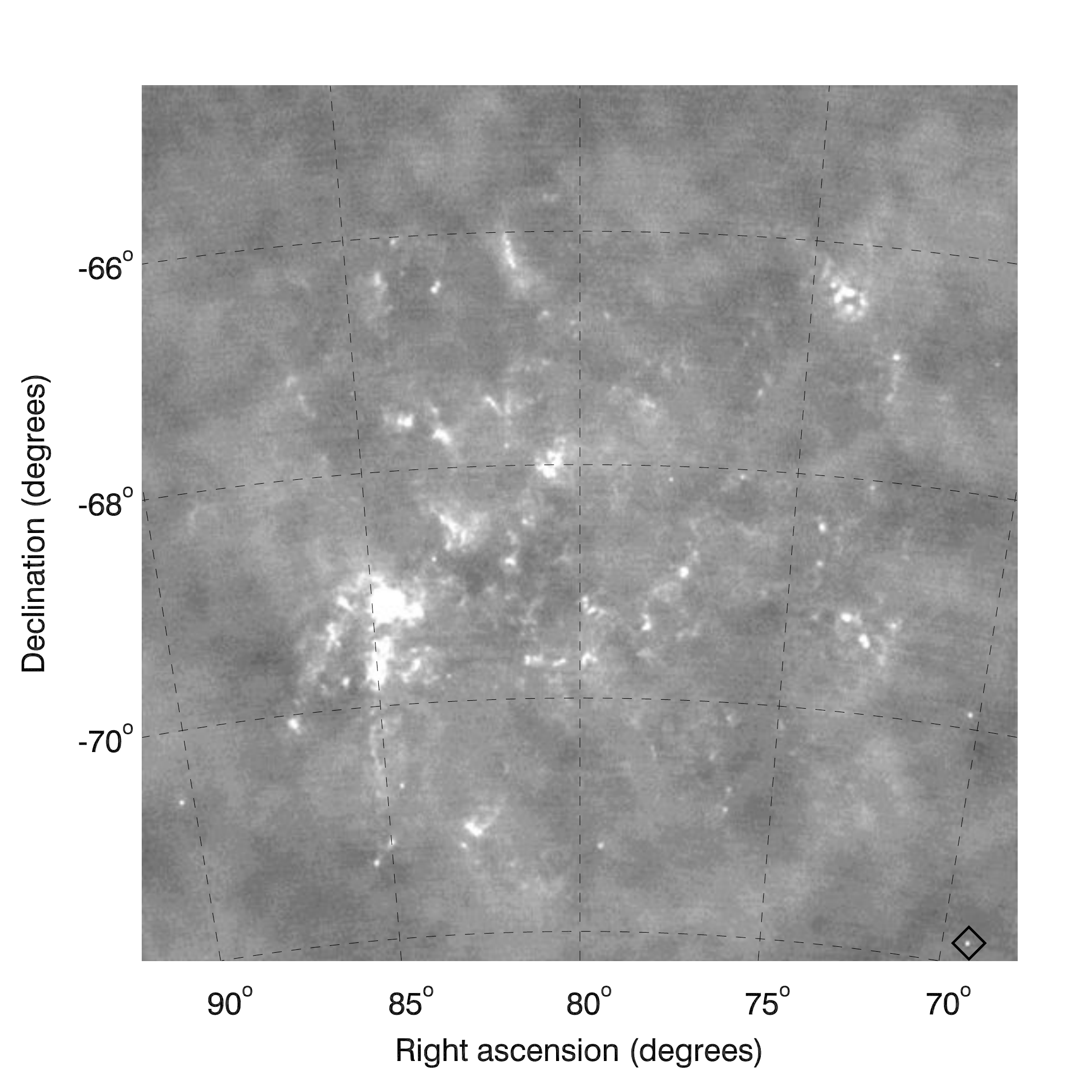}}
\subfigure[\planck+SPT map matched to the original \planck\ resolution ({\it Left Panel}); power spectrum
of the \planck-only map and the \planck-matched combined map and their ratio ({\it Right panel})]{\includegraphics[width=2.5in]{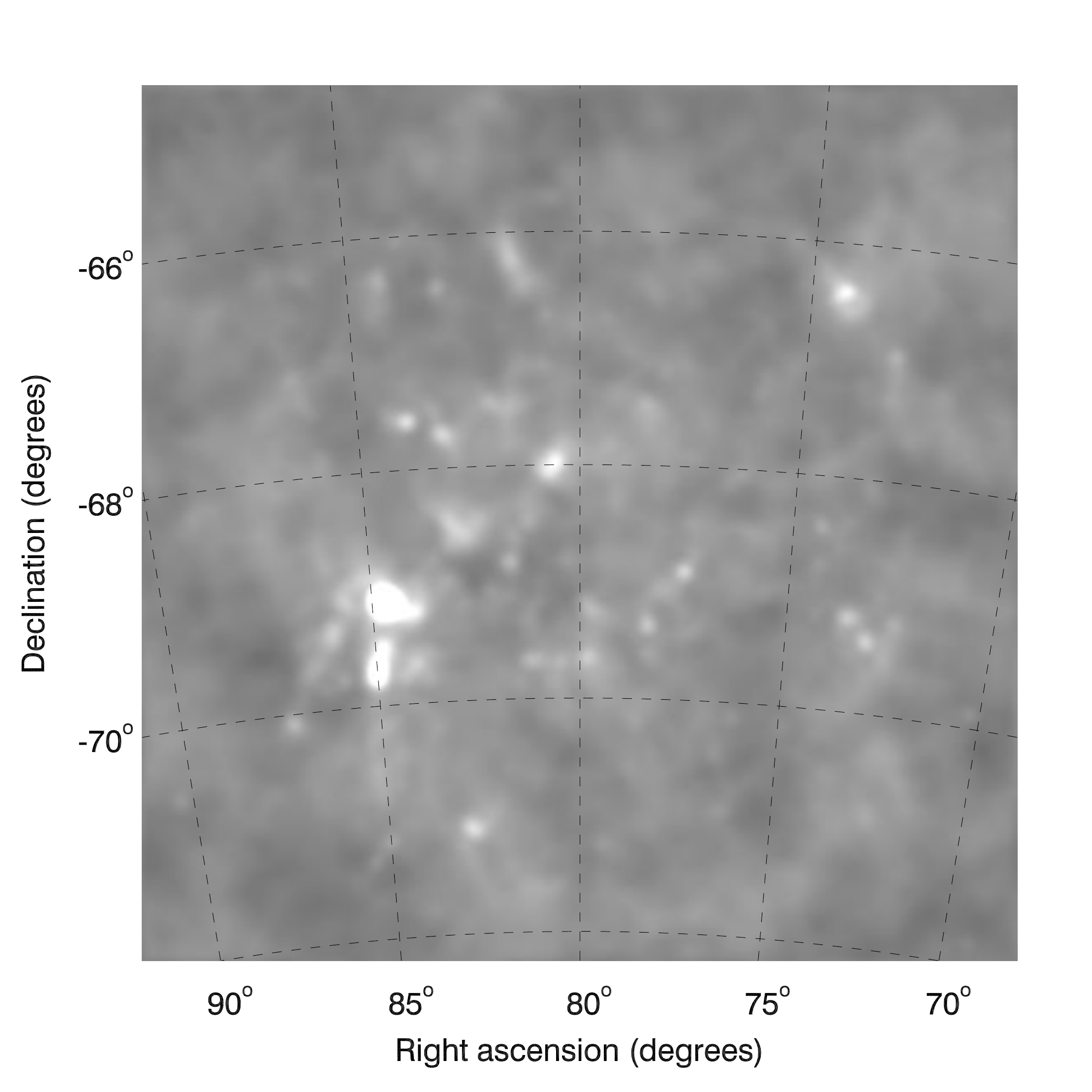}
\includegraphics[width=2.8in, trim = 0in -0.5in 0in 0in]{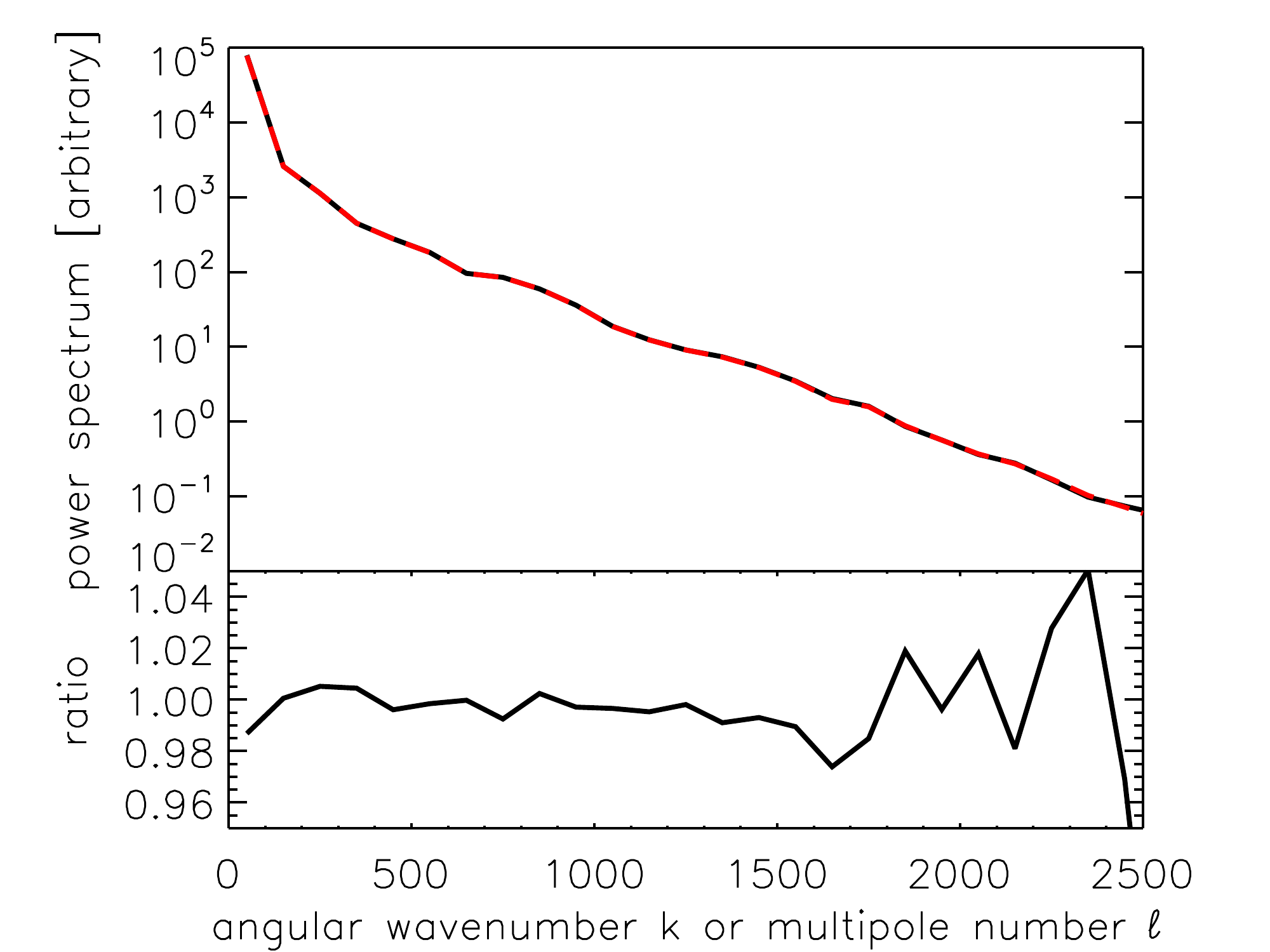}}
\caption{An illustration of the process of combining SPT and \planck\ data on the LMC field, and
a demonstration that the combined map is dominated at large scales by the \planck\ information, as expected.
\textbf{Top row:} The \planck\ 2.1~mm map, projected from the original HEALPix format
onto the 7.5-by-7.5-degree ZEA grid ({\it Left Panel}); 
the SPT 2.1~mm map on the native 7.5-by-7.5-degree ZEA grid ({\it Center Panel}); 
and the combined 2.1~mm map at 1.5~arcmin resolution ({\it Right Panel}). 
\textbf{Bottom row:} The combined 2.1~mm map convolved with a kernel equivalent to the
ratio of the 2.1~mm \planck\ beam to a 1.5-arcmin Gaussian to match the resolution of the 
original \planck\ map ({\it Left Panel}); and the power spectrum of the maps in the 
upper left (black solid line) and lower left (red dashed line) panels, as well as their ratio ({\it Right Panel},
see Section~\ref{sec:checks} for details on the power spectrum calculation).
The visual agreement between the maps in the upper left and lower left panels 
demonstrates that the
combined map reverts to the original \planck\ map when the higher-resolution SPT information
is smoothed away. This is shown quantitatively in the power spectrum ratio, which 
is within 5\% of unity at all scales at which there is significant information in the 
\planck-resolution maps.
The negative shadowing around regions of strong emission in the SPT-only
map is due to the common-mode subtraction (see Section~\ref{sec:sptxfer}) 
and goes away when the information in these angular
modes is added from the \planck\ map.
For these images and this calculation, we have used the combined map constructed assuming 
an emission spectrum $I(\lambda) \propto \lambda^{-2}$.
The black diamond in the upper-right panel indicates the position of PKS~0437-719, the
extragalactic source used for the resolution test shown in Figure~\ref{fig:reso}.
}
\label{fig:reverttoplanck}
\end{centering}
\end{figure*}

\begin{figure*}
\includegraphics[width=7in]{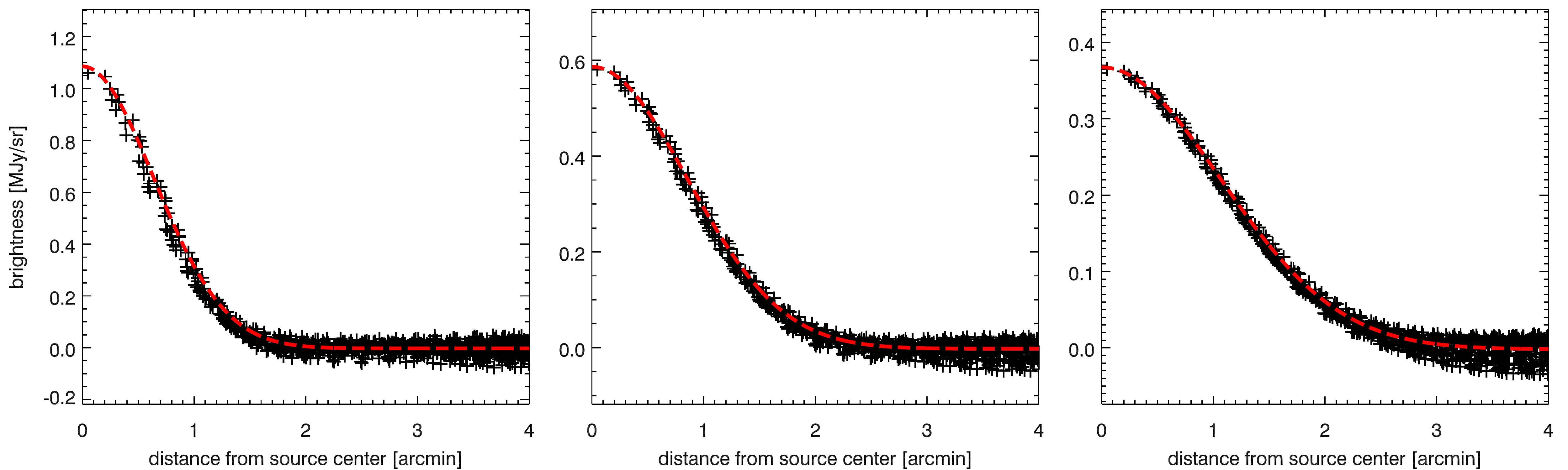}
\caption{Pixel values vs.~angular distance from the radio source PKS~0437-719, the brightest
background source in the LMC field. 
(The location of PKS~0437-719 is indicated by a black diamond in the upper-right panel of
Figure~\ref{fig:reverttoplanck}.)
This source is expected to be point-like at SPT resolution.
Pixel values are extracted from the 2.1~mm combined
SPT-\planck\ maps that are expected to have angular response functions equal to 2d Gaussians
with FWHM of 1.5 (\textbf{left panel}), 2.0 (\textbf{center panel}), and 2.5 (\textbf{right panel}) arcmin.
In each panel, the extracted pixel values are shown as black crosses, and the expected angular
response function (with amplitude taken from a fit of the map cutout to a 2d Gaussian) is shown
as a dashed red line. The measured shape of this radio source in the maps is consistent with
the expected angular response function.
}
\label{fig:reso}
\end{figure*}

\section{Results}
\label{sec:results}
The primary result of this work consists of two sets of 40 maps (one set each for the LMC and SMC
fields). 
These maps are 1800-by-1800 pixels and 1200-by-1200 
pixels---or 7.5-by-7.5 degrees and 5-by-5 degrees---for the LMC and SMC fields, respectively.
Each set of 40 maps consists of eight maps each created assuming one of five emission
underlying spectra (power-law emission $I(\lambda) \propto \lambda^{-\alpha}$ with spectral index 
$\alpha = \{-1.0,0.0,1.0,2.0,3.0\}$). For each value of spectral index, the eight maps 
consist of three maps of combined SPT and \planck\ data at 1.4~mm (one each at 
resolutions of 1.5, 2.0, and 2.5~arcmin), three maps of combined SPT and \planck\ data at 
2.1~mm (one each at resolutions of 1.5, 2.0, and 2.5~arcmin), and two maps of combined SPT 
and \planck\ data at  3.0~mm (one each at resolutions of 2.0, and 2.5~arcmin). 

In this Section,
we perform some simple tests to verify certain assumptions
or expectations about the maps, in particular their resolution and
their fidelity to the original \planck\ data. These tests and the results are discussed 
in Section \ref{sec:checks}. In Section \ref{sec:cutouts}, we show images from a
selection of maps, centered on
certain features of interest in the LMC and SMC, and discuss the properties
of the maps evident from these images. Finally, we 
discuss the instrumental and astrophysical noise properties of the maps
in Section \ref{sec:mapnoise}.

\subsection{Combined Map Checks}
\label{sec:checks}
In this section, we perform three checks on the combined SPT+\planck\ maps and the process used to construct
them. First, we use simulated observations to calculate the effect of ignoring the 
thin stripe of low-$k_x$, high-$k_y$ modes removed by the SPT scan-direction 
filtering and not replaced with Planck data (see Section \ref{sec:sptdecon} for details).
Next we verify the fidelity of the final, combined maps by comparing them with the original \planck\ data. 
Because of the nature of the
SPT and \planck\ data, in particular the filtering of large
angular scales (low-$k$ Fourier modes) from the SPT data (see Section \ref{sec:weights}
for details), we expect the combined maps to be dominated on large scales (small values
of wavenumber $k$) by the information from the \planck\ maps, and we check that this
expectation is borne out. Finally, we confirm the expected angular response function of 
the final, combined, Gaussian-smoothed maps:
If our measurements 
of the \planck\ and SPT beams and of the SPT filter response are accurate, then we expect that 
the only angular response function in the final maps is the Gaussian smoothing
(except for adjustments of the overall mean intensity in the LMC or SMC region---see 
Section~\ref{sec:galforeg} for details).

To estimate the bias that results from ignoring the small fraction of Fourier modes removed
from the SPT maps and not replaced by \planck\ data, we create simulated maps with
the same filtering as the real SPT maps and combine them with simulated \planck\ maps
of the same mock skies. We then perform aperture photometry on the simulated combined
SPT+\planck\ maps and compare the results to aperture photometry on the true, underlying
mock skies. We create many mock skies with features on different angular scales, and we
perform aperture photometry using many different aperture radii. For features on scales of
1 to 10 arcminutes and aperture radii in the same range, we find a typical bias of $<2\%$ and
a maximum bias of $<4\%$ resulting from the missing low-$k_x$, high-$k_y$ modes.

We verify the fidelity to the original \planck\ maps in two
ways. First, in Figure~\ref{fig:reverttoplanck} we show four versions of the 2.1~mm, $\alpha=2.0$ map of the
LMC field: 1) the \planck\ data (not beam-matched to SPT) directly projected onto the final ZEA grid; 
2) the filtered SPT data projected onto the final ZEA grid; 3) the 1.5-arcmin-FWHM version of the final
map; 4) the map in (3) convolved with the difference between a 1.5-arcmin-FWHM Gaussian and the 
\planck\ 2.1~mm beam. There is strong visual agreement between the original \planck\ map and the
final maps smoothed to \planck\ resolution. 
To make this more quantitative, we calculate the power spectrum of each of these maps
and plot these and the ratio between them in Figure~\ref{fig:reverttoplanck}. To avoid noise bias
in the power spectrum, we create two versions of each map using only half the SPT or \planck\ data
and calculate a cross-spectrum between the two half-depth maps. We mask the region around
30 Doradus before computing the power spectrum, as it otherwise dominates the power on all scales.
As shown in Figure~\ref{fig:reverttoplanck}, the power spectrum calcluated from these two 
maps agrees to better than 5\% at all scales on which there is significant power in the maps.
These two results support the idea that
these maps are dominated by \planck\ information on scales larger than the \planck\ beam. 

Second, we perform aperture photometry on the LMC field and SMC field maps, centered at R.A.=$78.88^\circ$,
declination=$-68.50^\circ$ and R.A.=$16.07^\circ$, declination=$-72.86^\circ$, respectively, and 
in apertures of radius $0.5^\circ$, $1^\circ$, and $2^\circ$. We then perform aperture photometry
on the \planck\ maps in their original HEALPix format, and we compare the resulting flux values.
The fluxes from aperture photometry on the maps presented here agree with the results of aperture
photometry on the original \planck\ maps to no worse than 2\% in all combinations of wavelength
band, field, and aperture size. These results are consistent with the aperture photometry on
simulated maps.

Finally, we expect these maps
to be nearly unbiased estimates of the sky brightness at all angular scales; the only response
function expected is the 1.5, 2.0, or 2.5-arcmin FWHM Gaussian smoothing
kernel and the small strip of modes missing at low $k_x$ and high $k_y$.
We confirm this expectation by taking a cutout of the 2.1~mm ($\alpha=2.0$) map of each resolution around the brightest
background point source in the LMC field, the radio source PKS~0437-719
(which is expected to be point-like at SPT resolution, \citealt{healey07}). 
The location of this source is indicated by a black diamond in the upper-right panel of
Figure~\ref{fig:reverttoplanck}.
The brightness of the map at each resolution 
as a function of distance to this source is plotted in Figure \ref{fig:reso}. Overplotted
in each case is the expected response to a point source in that map, namely a 1.5, 2.0, or 2.5-arcmin
FWHM Gaussian. The measured response is consistent by eye with the expected response. If
we fit the map cutouts to a model of a two-dimensional Gaussian, the geometric mean of the best-fit FWHM
along the two axes are 1.45, 1.98, and 2.51 arcmin at the three resolutions
(within $\sim$3\% of the expected FWHM of 1.5, 2.0 and 2.5~arcmin).

\begin{figure*}
\begin{centering}
\subfigure[LMC, molecular ridge below 30 Doradus, 1.4~mm, 1.5-arcmin resolution]{ 
 \includegraphics[width=6in]{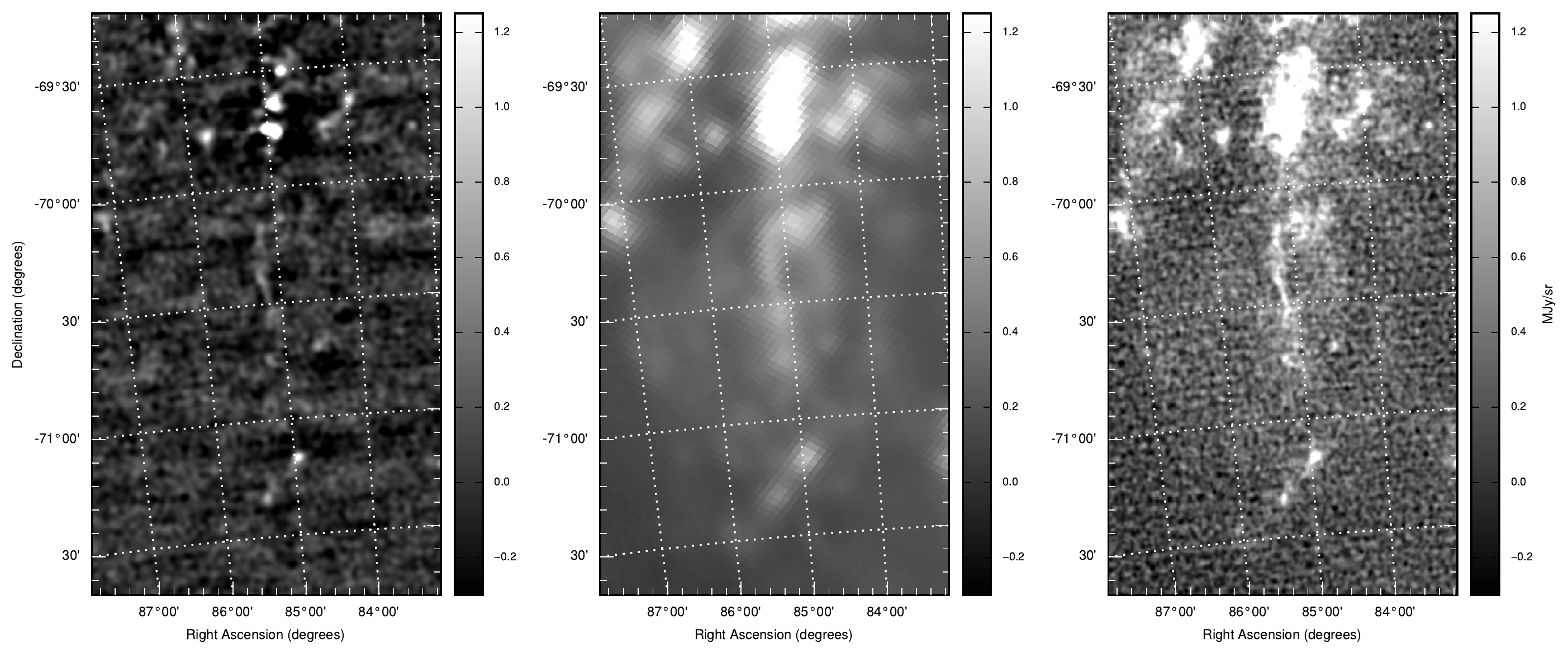}}
\subfigure[LMC, molecular ridge below 30 Doradus, 2.1~mm, 1.5-arcmin resolution]{ 
 \includegraphics[width=6in]{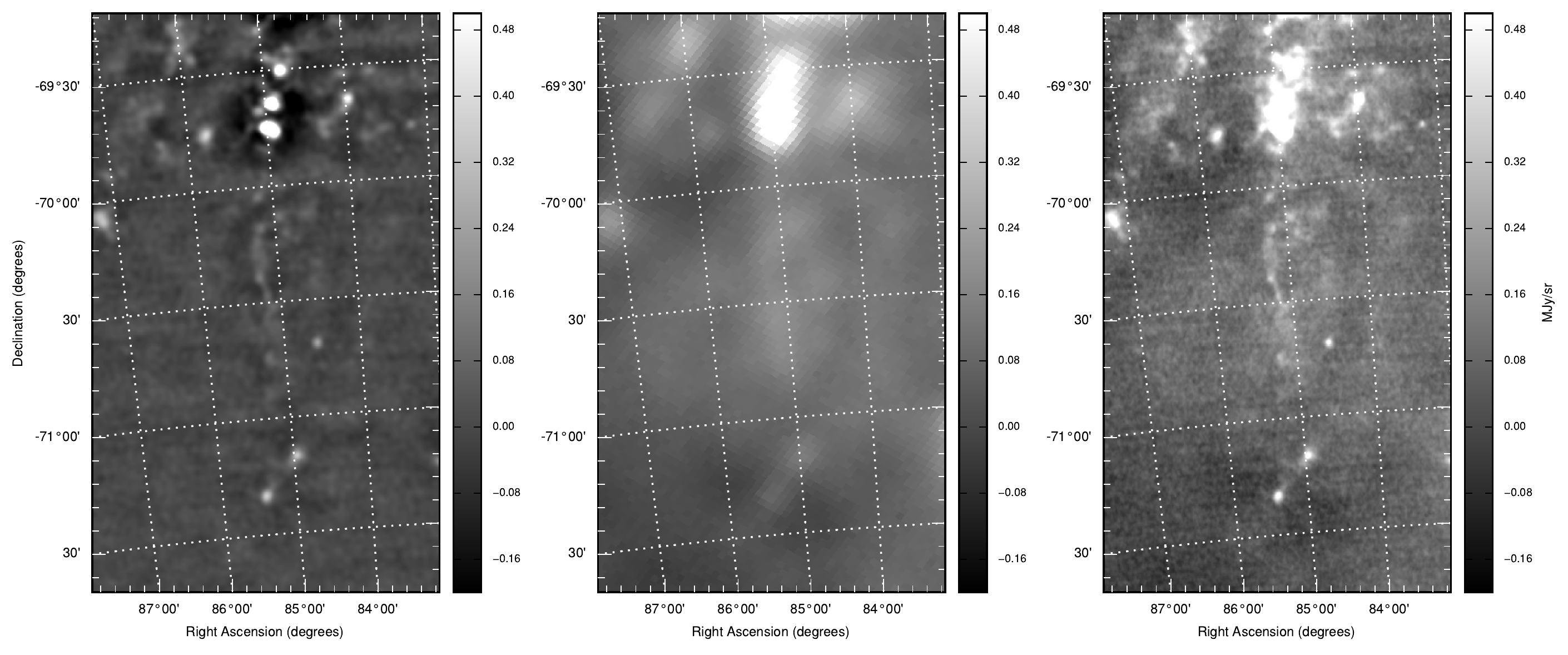}}
\subfigure[LMC, molecular ridge below 30 Doradus, 3.0~mm, 1.8-arcmin resolution]{ 
 \includegraphics[width=6in]{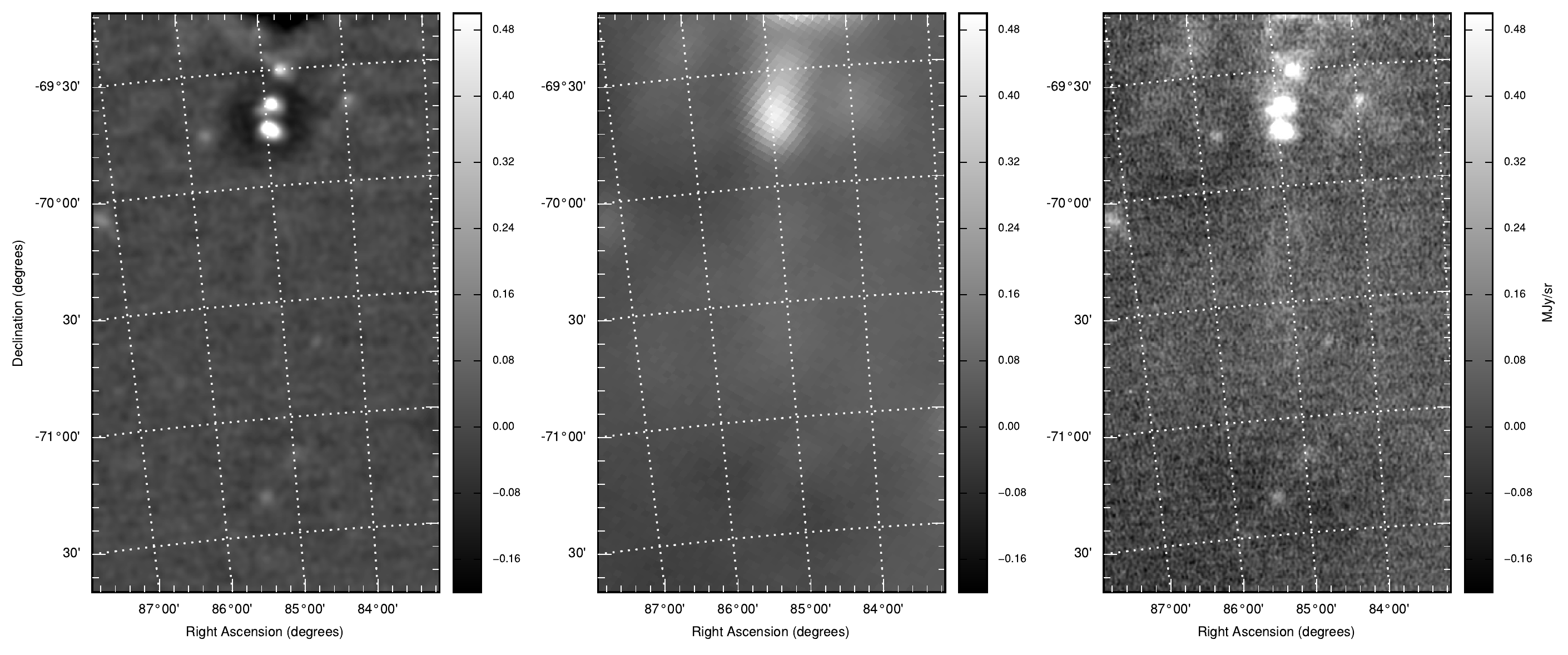}}
\caption{
\label{fig:zoommaps1} 
Combined SPT-\planck\ maps of the molecular ridge below 30 Doradus in the LMC.
This subfield of the LMC is chosen to emphasize the gain in resolution between \planck\
alone and \planck+SPT. The field in shown in SPT data alone ({\it Left}), in \planck\ data 
alone ({\it Center}), and combined ({\it Right}) in all three wavelength bands.
The units of the maps are MJy~sr$^{-1}$, and
the SPT and \planck\ data that make up the maps have been converted from 
CMB fluctuation temperature to specific intensity assuming an underlying emission
spectrum $I(\lambda) \propto \lambda^{-2}$.}
\end{centering}
\end{figure*}

\begin{figure*}
\begin{centering}
\subfigure[LMC, N11, 1.4~mm, 1.5-arcmin resolution]{ 
 \includegraphics[width=7in]{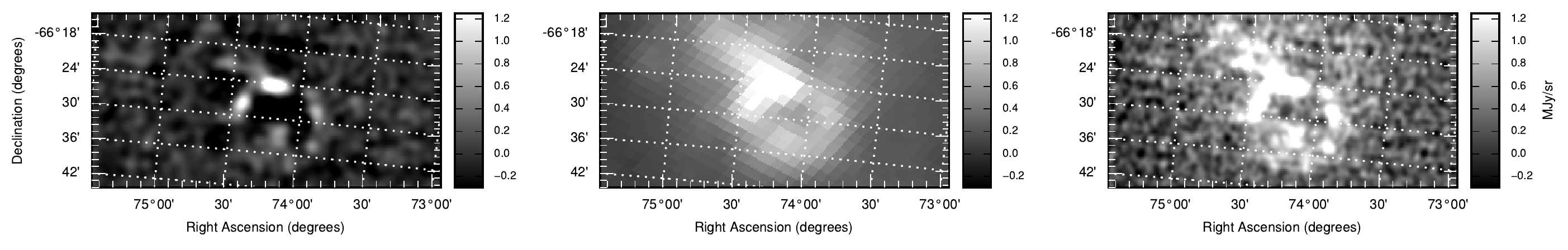}}
\subfigure[LMC, N11, 2.1~mm, 1.5-arcmin resolution]{ 
 \includegraphics[width=7in]{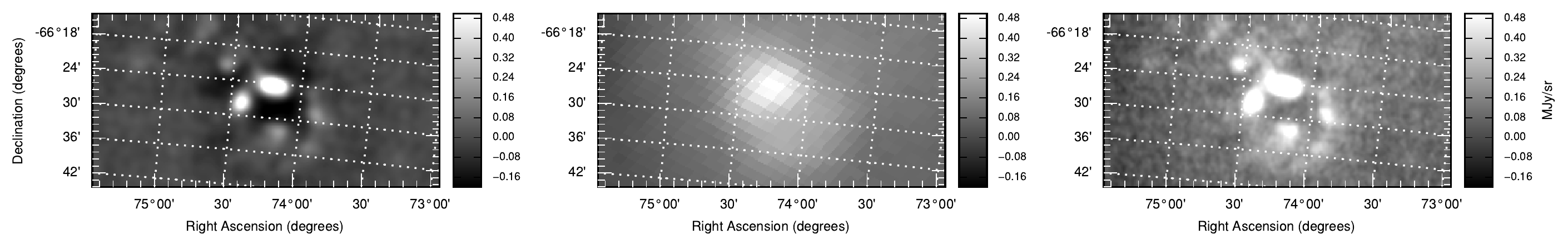}}
\subfigure[LMC, N11, 3.0~mm, 1.8-arcmin resolution]{ 
 \includegraphics[width=7in]{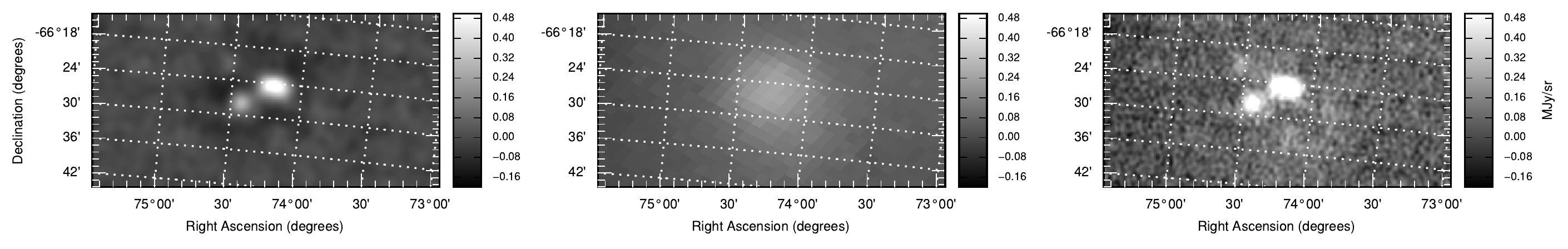}}
\caption{
\label{fig:zoommaps2} 
Combined SPT-\planck\ maps of the star-forming region N11 in the LMC.
This subfield of the LMC is chosen to emphasize the gain in resolution between \planck\
alone and \planck+SPT. The field in shown in SPT data alone ({\it Left}), in \planck\ data 
alone ({\it Center}), and combined ({\it Right}) in all three wavelength bands.
The units of the maps are MJy~sr$^{-1}$, and
the SPT and \planck\ data that make up the maps have been converted from 
CMB fluctuation temperature to specific intensity assuming an underlying emission
spectrum $I(\lambda) \propto \lambda^{-2}$.}
\end{centering}
\end{figure*}

\begin{figure*}
\begin{centering}
\subfigure[SMC, 1.4~mm, 2.5-arcmin resolution]{ 
 \includegraphics[width=6in]{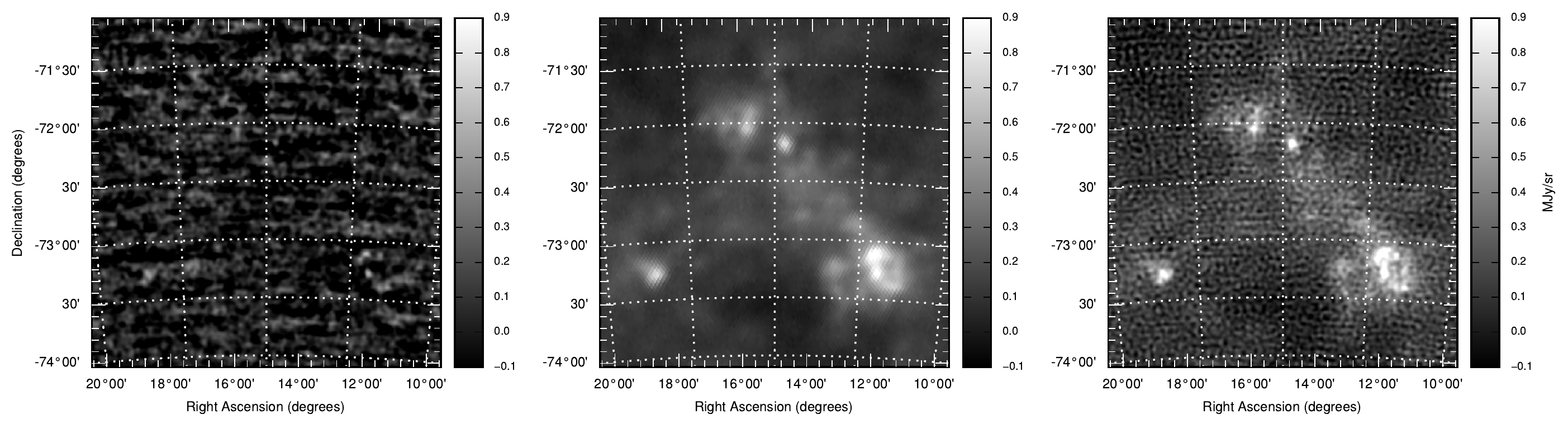}}
\subfigure[SMC, 2.1~mm, 2.5-arcmin resolution]{ 
 \includegraphics[width=6in]{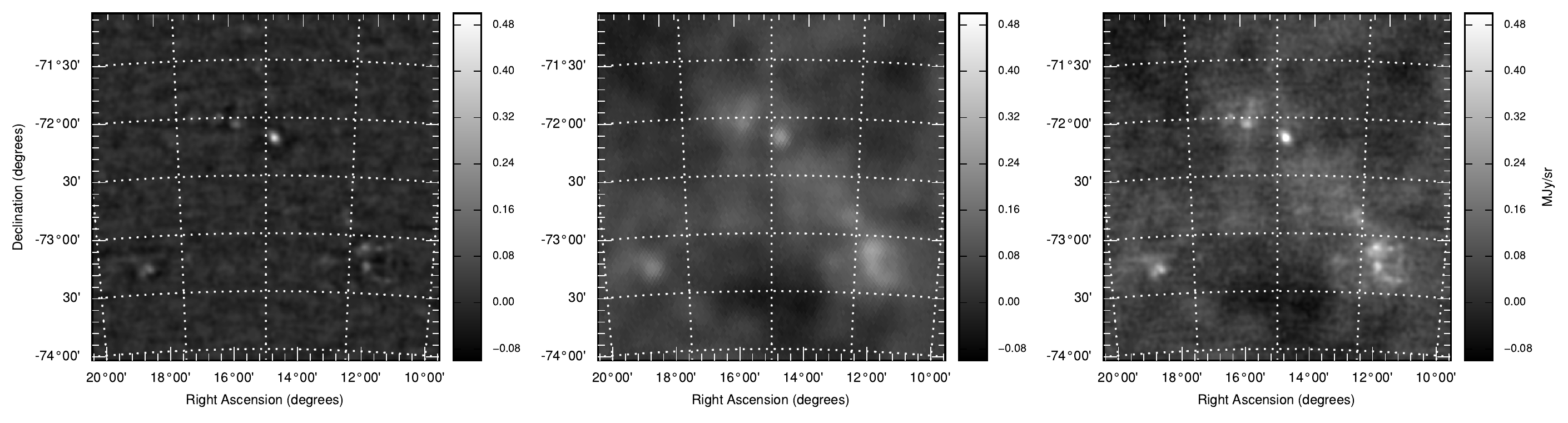}}
\subfigure[SMC, 3.1~mm, 2.5-arcmin resolution]{ 
 \includegraphics[width=6in]{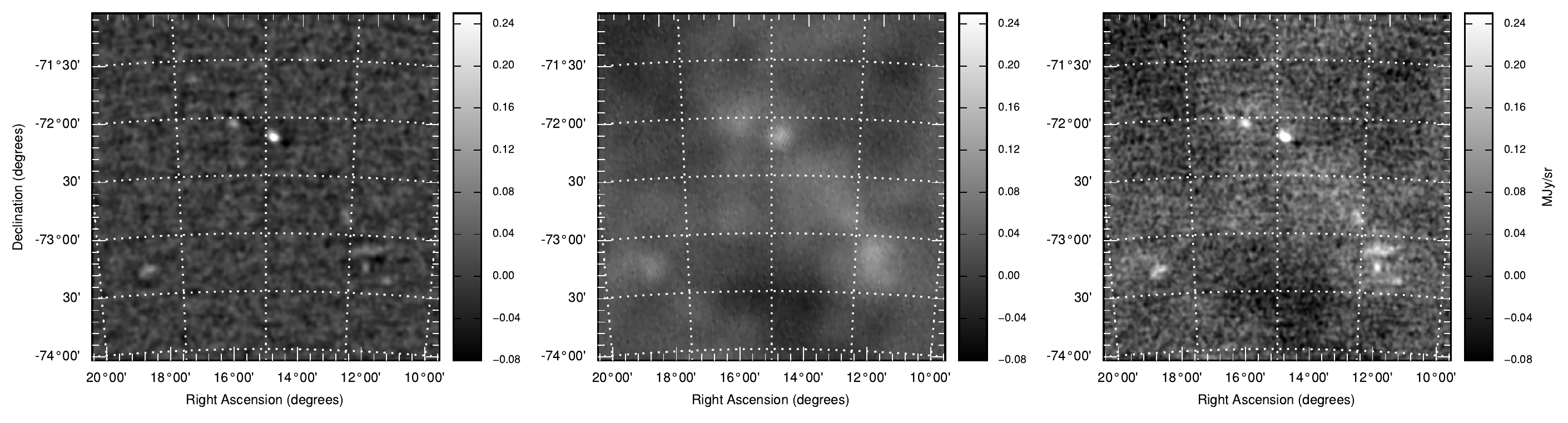}}
\caption{
\label{fig:zoommaps3} 
Combined SPT-\planck\ maps of the SMC.
The field in shown in SPT data alone ({\it Left}), in \planck\ data 
alone ({\it Center}), and combined ({\it Right}) in all three wavelength bands.
The units of the maps are MJy~sr$^{-1}$, and
the SPT and \planck\ data that make up the maps have been converted from 
CMB fluctuation temperature to specific intensity assuming an underlying emission
spectrum $I(\lambda) \propto \lambda^{-2}$.}
\end{centering}
\end{figure*}

\subsection{Selected Map Images}
\label{sec:cutouts}
In Figures~\ref{fig:zoommaps1}--\ref{fig:zoommaps3}, we show cutouts of a selection 
of \planck+SPT maps centered on features of interest in the LMC and SMC fields. 
Figure~\ref{fig:zoommaps1} shows the molecular ridge south of 30 Doradus in the LMC
(e.g., \citealt{ott08}), 
Figure~\ref{fig:zoommaps2} shows the star-forming region N11 in the LMC,
and Figure~\ref{fig:zoommaps3} shows a 2.5-by-2.5 degree cutout of the full
5-by-5 degree SMC field.
In all cases, the maps shown use data converted from 
CMB fluctuation temperature to specific intensity assuming an underlying emission
spectrum $I(\lambda) \propto \lambda^{-2}$. The images of LMC regions (Figures~\ref{fig:zoommaps1}
and \ref{fig:zoommaps2}) use maps at 1.5-arcmin resolution for 1.4 and 2.1~mm and maps 
at 1.8-arcmin resolution for 3.0~mm. The image of the SMC (Figure~\ref{fig:zoommaps3})
uses 2.5-arcmin resolution maps at all wavelengths.

In all of the combined maps, it is clear
there is ample arcminute-scale structure in the millimeter-wave
emission from Magellanic Clouds (particularly the LMC) and that this structure
is qualitatively similar in the three wavelength bands used in this work. Comparing
the \planck-only map in each figure to the combined map
demonstrates the value of adding the higher-resolution SPT data in 
elucidating this small-scale structure. The one possible exception to this is 
the 1.4~mm map of the SMC, in which the noise in the SPT map is high enough
that the SPT data contributes comparatively little to the combined map. No obvious
artifacts are visible in any of these images.

\subsection{Noise Properties of the Combined Maps}
\label{sec:mapnoise}
In this section, we discuss the noise properties of the combined SPT-\planck\ maps. For the 
purposes of measuring emission from the Magellanic Clouds, we consider astronomical signal
from other sources to be noise. The only significant astronomical contribution to the noise budget
in this work is anisotropy in the CMB. We first discuss the contribution to the map noise from 
the two instruments, then we discuss the contribution from the CMB.

\subsubsection{Instrument Noise}
\label{sec:instnoise}
\begin{figure*}
\begin{centering}
\subfigure[LMC null map, 1.4~mm]{\label{fig:lmc1p4smooth1p5} 
  \includegraphics[width=2.25in]{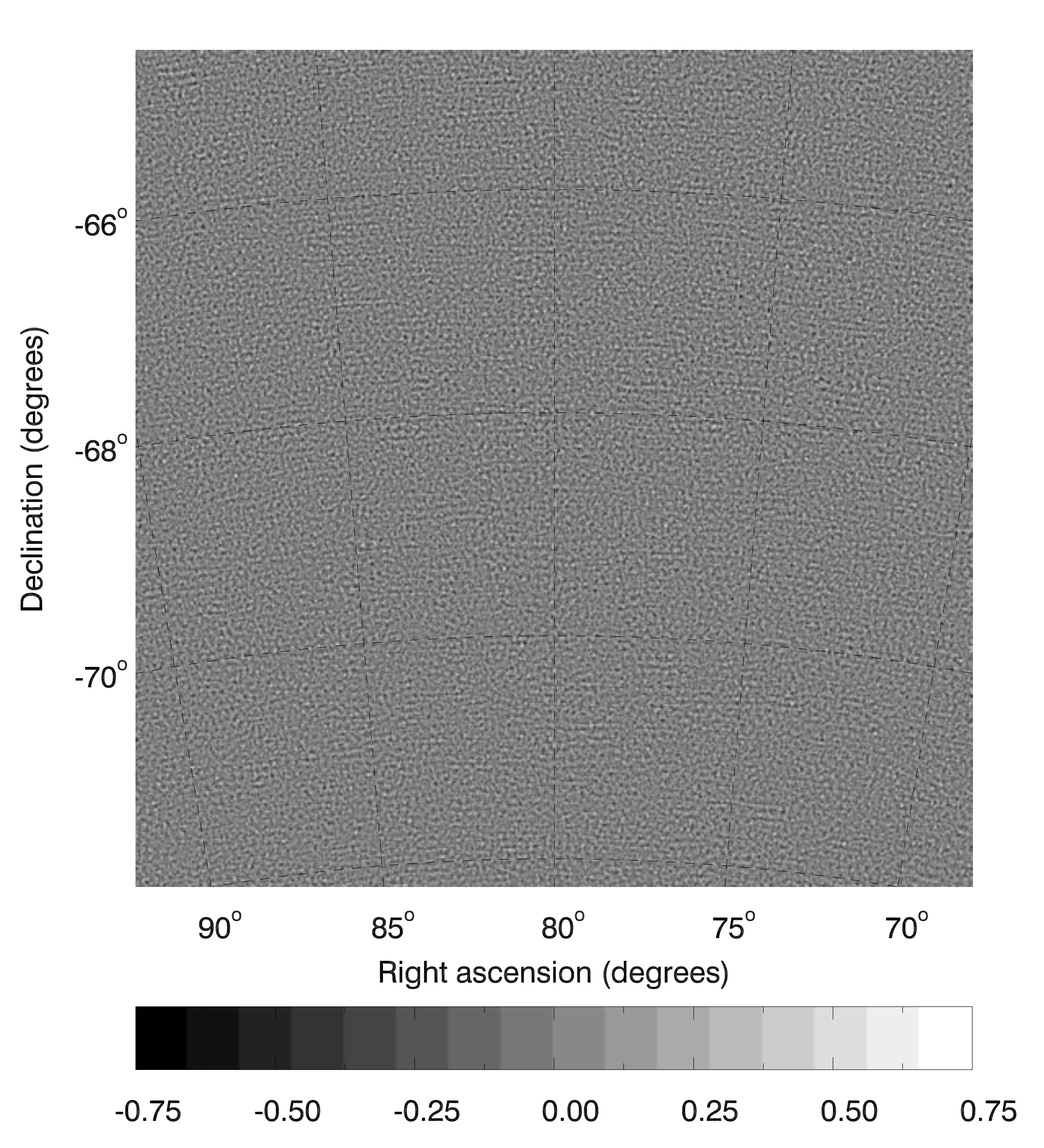}}
\subfigure[LMC null map, 2.1~mm]{\label{fig:lmc2p1smooth1p5} 
  \includegraphics[width=2.25in]{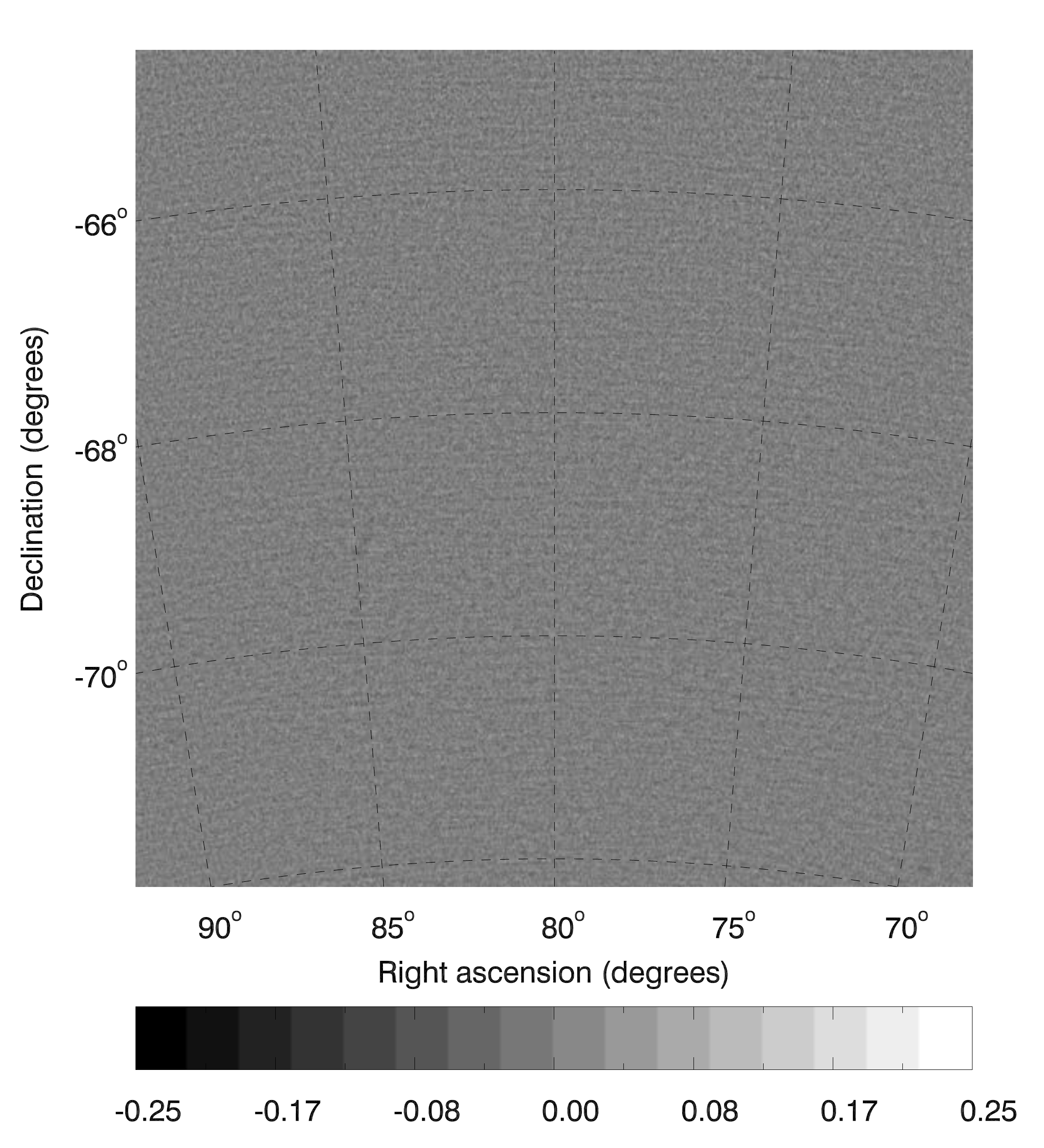}}
\subfigure[LMC null map, 3.0~mm]{\label{fig:lmc3p0smooth2p0} 
  \includegraphics[width=2.25in]{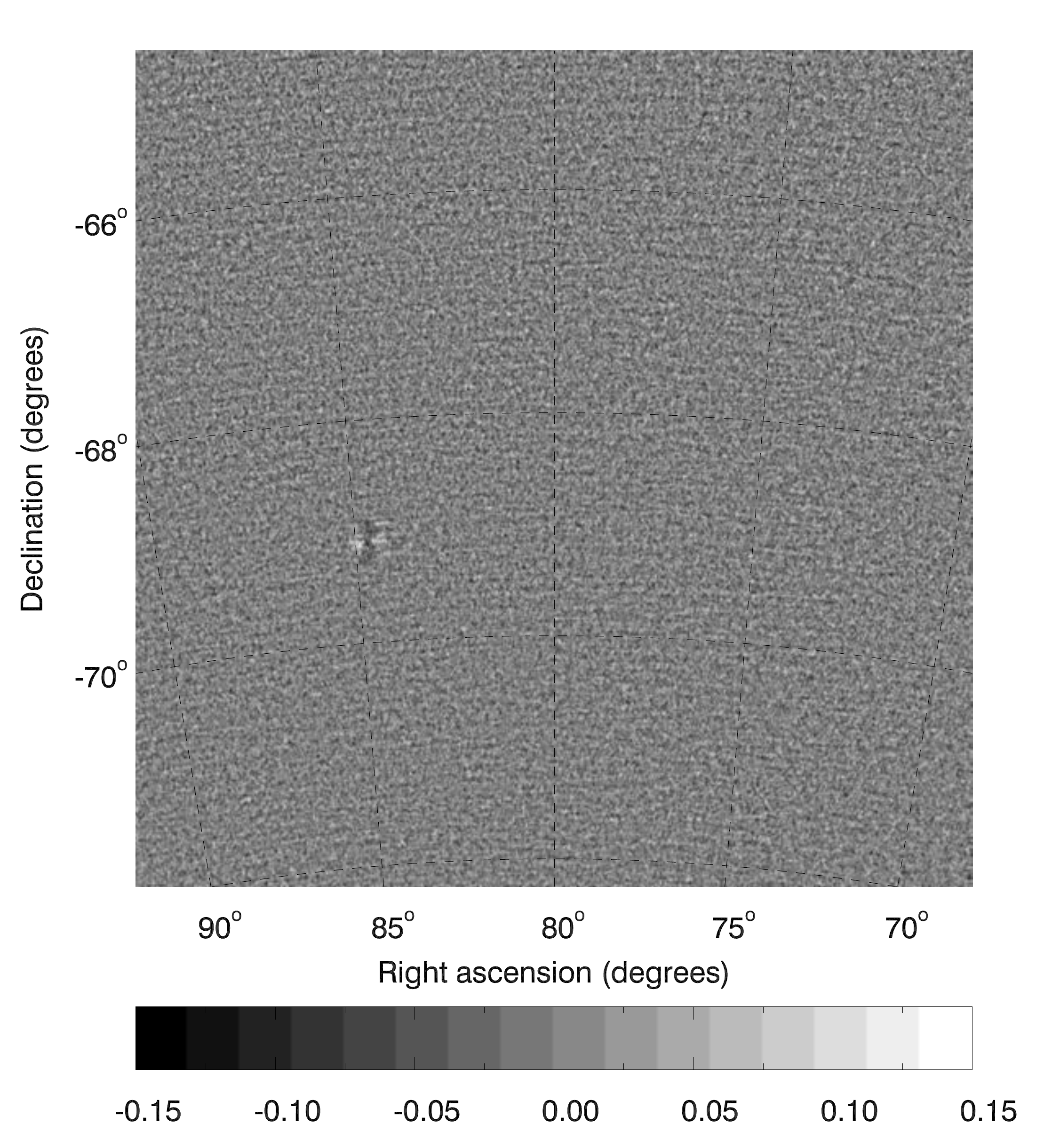}}
\caption{
\label{fig:lmcnullmaps} 
Combined SPT-\planck\ null maps of the 7.5-by-7.5-degree LMC field in three wavelength
bands. Null maps are created by subtracting one half of the data from the other half
and dividing by two. The 1.4~mm and 2.1~mm are at 1.5-arcmin resolution; the 3.0~mm
map is at 2.0-arcmin resolution. The units of the maps are MJy~sr$^{-1}$.
The localized feature at roughly [$85^\circ$,$-69^\circ$] in the 3.0~mm map is 
a residual of the bright LMC source 30 Doradus (see Sections \ref{sec:sptdata} and \ref{sec:planck}
for details).
For all images, we have used the combined null maps constructed assuming 
an emission spectrum $I(\lambda) \propto \lambda^{-2}$.
}
\end{centering}
\end{figure*}

Figure \ref{fig:lmcnullmaps} shows a real-space representation of the instrument noise contribution
to the LMC field map noise, as measured in the combined SPT-\planck\ null maps. The construction
of the null maps for each instrument is described in detail in Sections \ref{sec:sptdata} and \ref{sec:planck}.
We construct combined null maps by combining null maps from each instrument in the same way as the signal maps.
Instrument noise is also discernible without differencing away signal in some of the 
maps shown in Figures~\ref{fig:zoommaps1}--\ref{fig:zoommaps3}, particularly at 1.4 and 3.0~mm.

In the LMC null maps and in the cutout maps at 3.0~mm, the instrument noise
is most visible at the smoothing scale of the maps, as would be expected for noise that was
white before smoothing. However, it
is possible to discern an isotropic pattern of noise at a different scale in the 1.4~mm maps, 
particularly in the SMC.
This pattern is from pixel-scale \planck\ noise convolved with the ratio of the SPT and 
\planck\ 1.4~mm beams. This ratio is cut off at $k \sim 4000$, which imparts the particular angular
scale to the noise pattern. The reason this pattern is more visible in 1.4~mm than at the other 
wavelengths is the relative depths of the SPT and \planck\ maps: the ratio of SPT map noise to 
\planck\ map noise is significantly higher at 1.4~mm than in the other bands, thus the \planck\ map
contributes to the combination out to $k$ values at which the value of the \planck\ beam is quite small.

For practical purposes, the most important property of the map noise is the noise rms in map patches
of various size---i.e., the expected noise contribution to the uncertainty in the measurement of the 
brightness of a feature in the maps of a given angular size. In Figure \ref{fig:noisevrad}, we plot
the instrument noise contribution to the measurement of the flux of a feature as a function of the size of that 
feature. Specifically, this is the standard deviation of 400 measurements of the flux within a 
circular region of a given diameter in the null map, each measurement with a different (random) center.
We also show the contribution of the CMB to this uncertainty (discussed in the next section), and
we show both contributions in the case that the flux in an equal-area region around the circular region is 
subtracted (we refer to this as the compensated top hat flux). In Table \ref{tab:noisevrad}, we list
values of these contributions (for the compensated and uncompensated top hat fluxes) for several
values of the region diameter.

\subsubsection{Astrophysical Noise}
\label{sec:cmbnoise}
Comparing Figures \ref{fig:reverttoplanck} and \ref{fig:lmcnullmaps}, it is clear that there is a diffuse, 
large-scale contribution to the signal maps in the LMC field that is not in the null maps and hence 
not part of the instrument noise budget discussed in the previous section. 
The morphology of this signal is consistent with that of a random Gaussian field 
with the power spectrum of the CMB and 
inconsistent with the filamentary structure of high-latitude Galactic emission, indicating that
it is likely CMB anisotropy.
The signal is present in all three wavelength bands and in both the LMC and SMC maps, 
and its spectral shape in the three bands is consistent with that of CMB anisotropy and inconsistent
with that of thermal dust or synchrotron emission. This signal is not present in shorter-wavelength
\herschel-SPIRE maps (see, e.g., \citealt{meixner13} or Figure~\ref{fig:lmcmapswspire}), indicating it is not 
Galactic dust emission.
The rms temperature fluctuation in
the CMB, when measured in patches significantly smaller than a degree, is approximately 
$100 \ \mu \mathrm{K}$, corresponding to roughly 0.04, 0.035, and 0.02 MJy~sr$^{-1}$ in the three wavelength
bands used in this work. This is consistent with the amplitude of the diffuse background
in both the LMC and SMC maps.

\begin{figure*}
\subfigure[Uncompensated aperture]{
\includegraphics[width=3.5in]{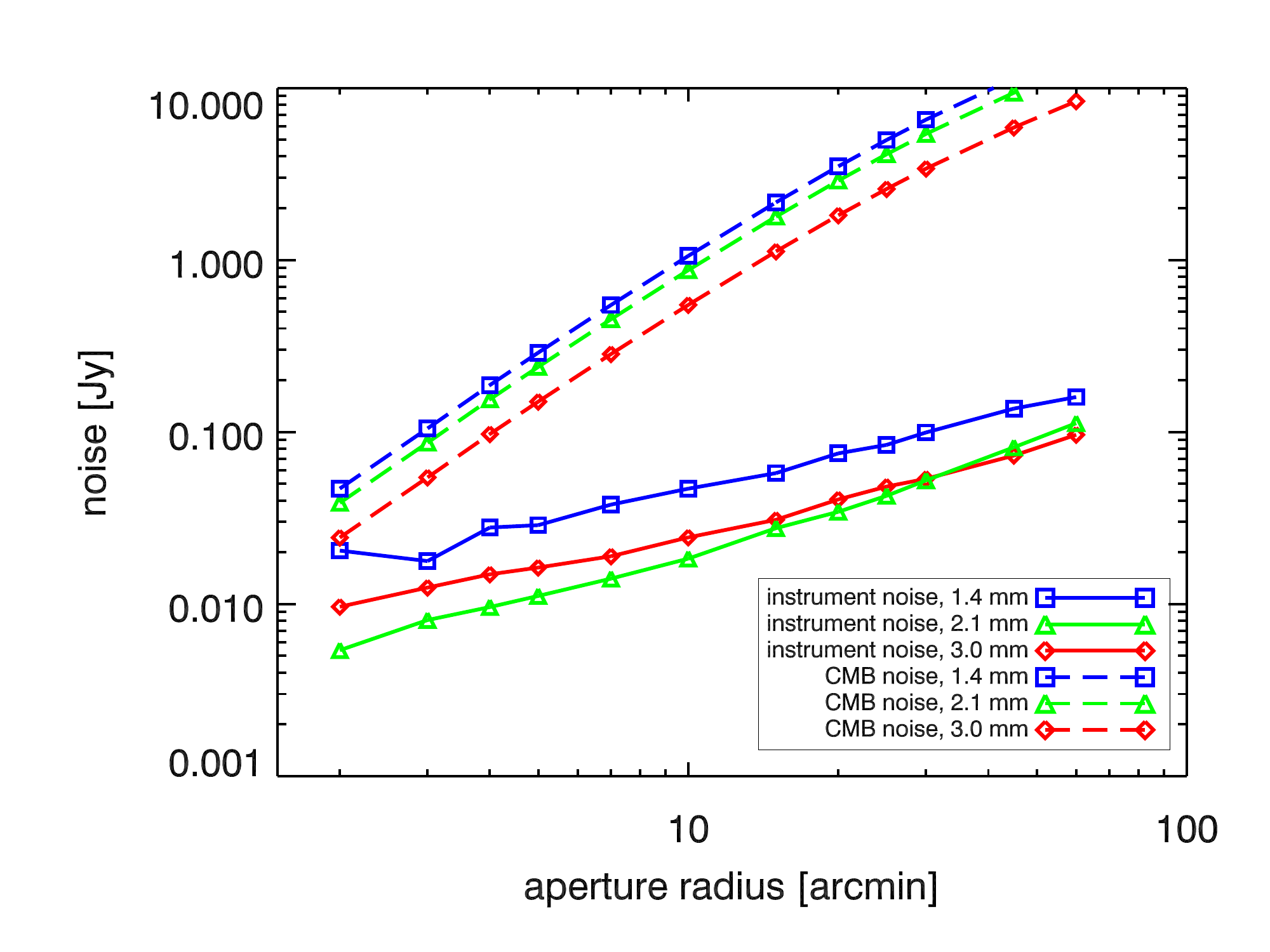}}
\subfigure[Compensated aperture]{
\includegraphics[width=3.5in]{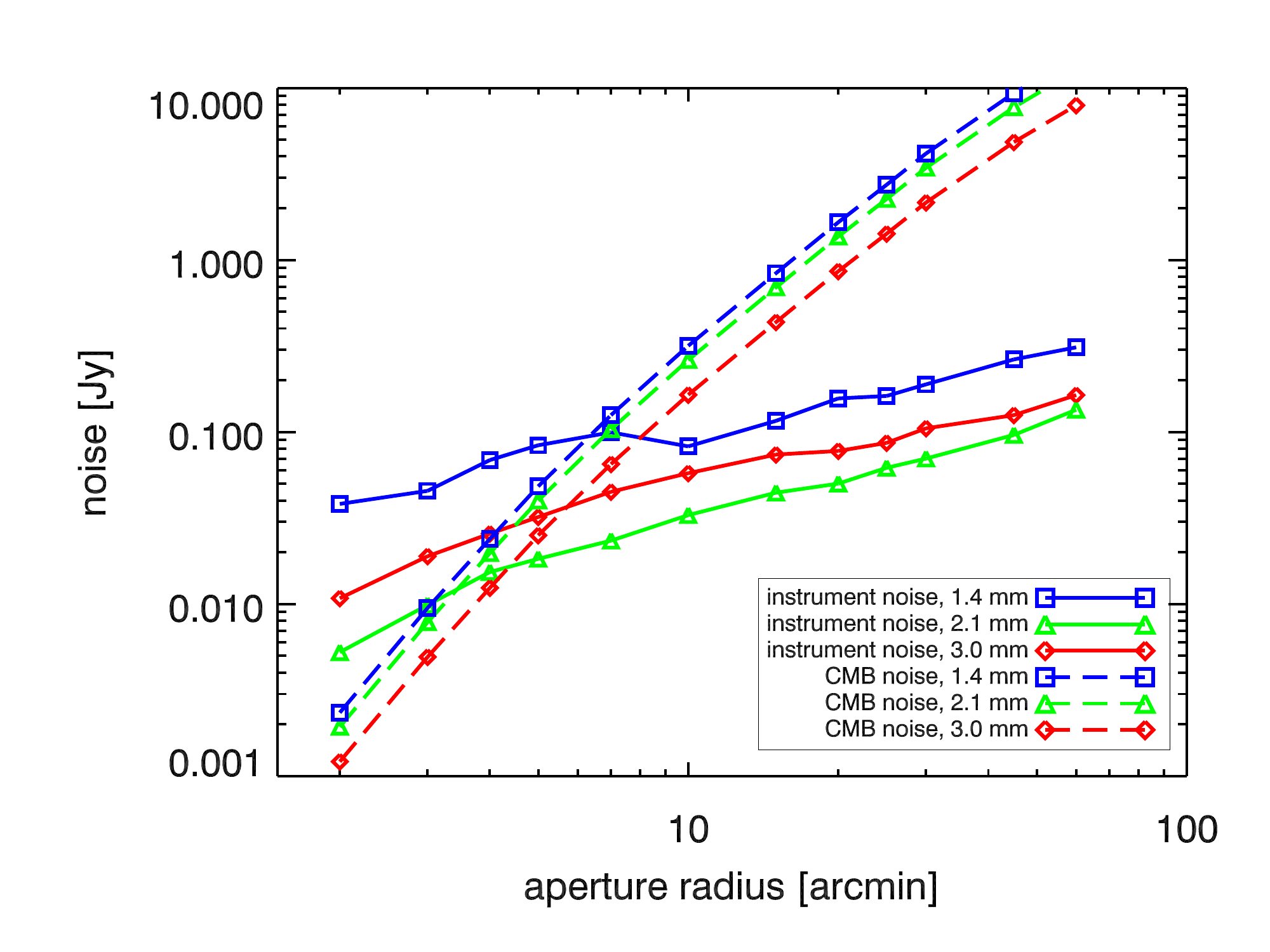}}
\caption{Expected contributions from instrument noise and CMB anisotropy 
to the uncertainty on a measurement of flux in the LMC field within an aperture 
as a function of aperture size. The instrument noise values in both panels are estimated from the SPT+\planck\
null maps at 2.0-arcmin resolution, and the CMB values are estimated from simulated 
CMB maps with 2.0-arcmin resolution. The behavior at 1.5 and 2.5-arcmin resolution is
qualitatively similar except at the smallest aperture, and the behavior in the SMC field
is nearly identical, but with slightly higher instrument noise values. Values at particular aperture sizes for
all map resolutions and both fields are shown in Table \ref{tab:noisevrad}. Values shown in the left panel
are calculated with no compensating negative region around the aperture; values shown in the
right panel are calculated with the flux from an equal-area region surrounding the aperture 
subtracted. For apertures smaller than $\sim 30$~arcmin, this compensation significantly reduces the contribution from the CMB at the expense
of increasing the instrument noise contribution (by roughly a factor of $\sqrt{2}$).
}
\label{fig:noisevrad}
\end{figure*}

Anisotropy in the CMB has a much different
behavior as a function of angular scale than the instrument noise in these maps: the CMB power
is highest on degree angular scales, while the instrument noise is more scale-independent. This 
leads to different relative contributions of the two noise sources at different scales, as shown in 
Figure \ref{fig:noisevrad} and Table \ref{tab:noisevrad}. The CMB contribution in this figure and plot
are calculated by creating many simulated CMB skies, converting from CMB temperature fluctuation
to specific intensity or brightness in each wavelength band, and performing aperture photometry on
the simulated maps. As with the instrument noise contribution, the standard deviation of the 
flux measured in many apertures of a given size around random centers is reported as the expected
noise contribution.

The different angular scale dependence of the CMB contribution to the noise means that, on scales of less than about
a degree, the noise contribution from the CMB can be suppressed by a simple spatial filtering operation
such as subtracting an equal-area region around an aperture in which one wishes to measure a flux.
Of course, if there is significant flux in the compensating aperture from Magellanic Cloud features, 
this operation will not simply difference away the CMB but also bias the aperture measurement.
We report expected CMB noise levels (Table \ref{tab:noisevrad}) for making flux measurements performed 
with such a compensated top-hat filter, but we caution that such a filter is most appropriate for isolated 
structures, rather than for crowded fields.

\section{Comparison with \herschel-SPIRE Maps of the LMC}
\label{sec:herschel}
As mentioned in Section~\ref{sec:intro}, \citet{meixner13} have produced maps from the 
\herschel\ HERITAGE survey of the Magellanic Clouds. Of all publicly available data on the
Magellanic Clouds, these maps are
closest in wavelength and resolution to the maps produced here, and comparing the two 
sets of maps provides both a visual check on the maps produced in this work and insight
into the emission processes at work in the Magellanic Clouds. We focus on the LMC here, because
the signal-to-noise in the SPT-\planck\ maps is higher than in the SMC. We further focus on 
the HERITAGE maps from the Spectral and Photometric Imaging Receiver (SPIRE) instrument
rather than from the Photodetector Array Camera and Spectrometer (PACS) instrument, because
the SPIRE bands are closer in wavelength to the SPT-\planck\ bands used here.

\begin{figure*}
\begin{centering}
\subfigure[LMC, 500~$\mu$m, 2.0-arcmin resolution]{
  \includegraphics[width=3.25in]{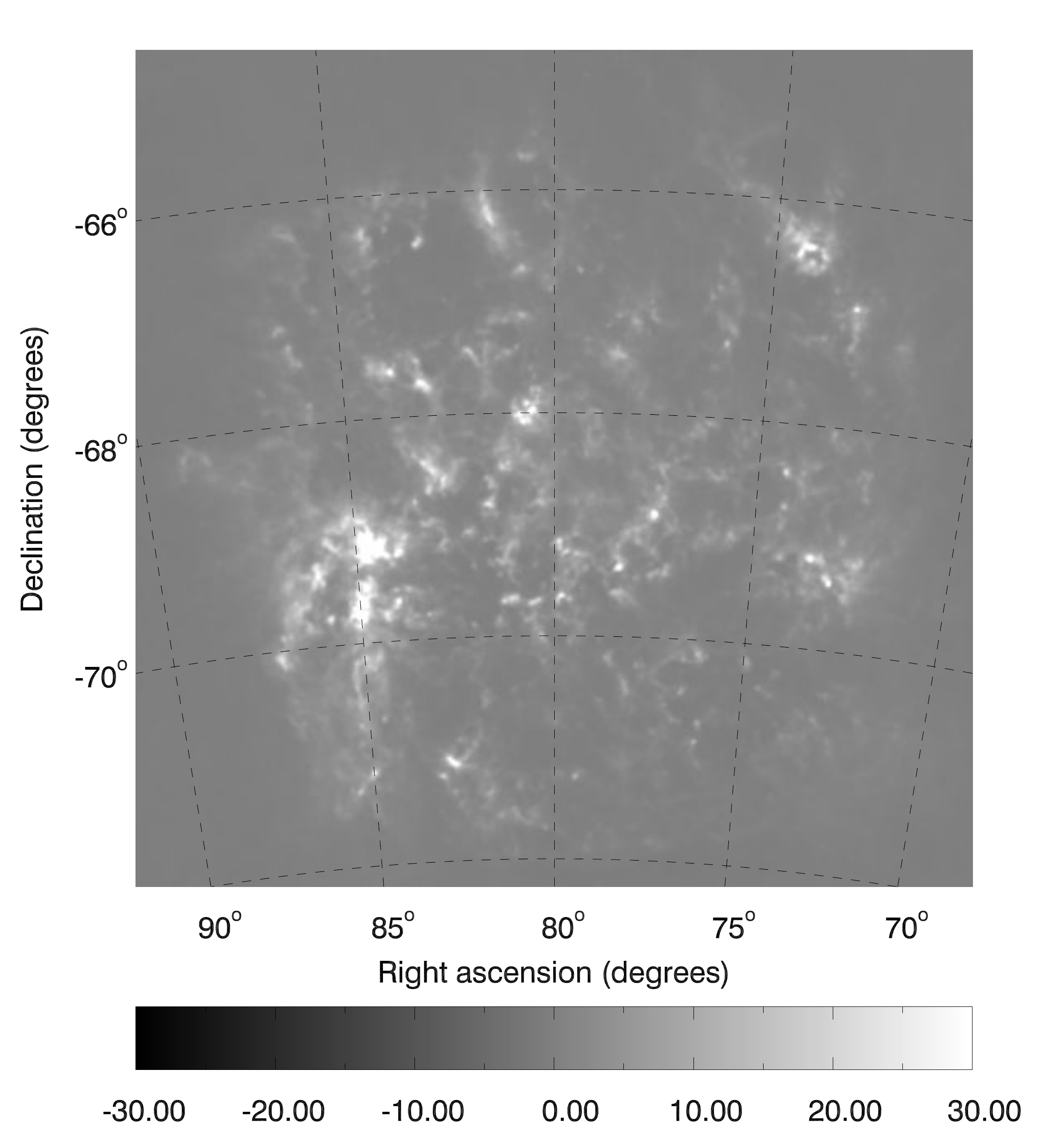}}
\subfigure[LMC, 2.1~mm, 2.0-arcmin resolution]{
  \includegraphics[width=3.25in]{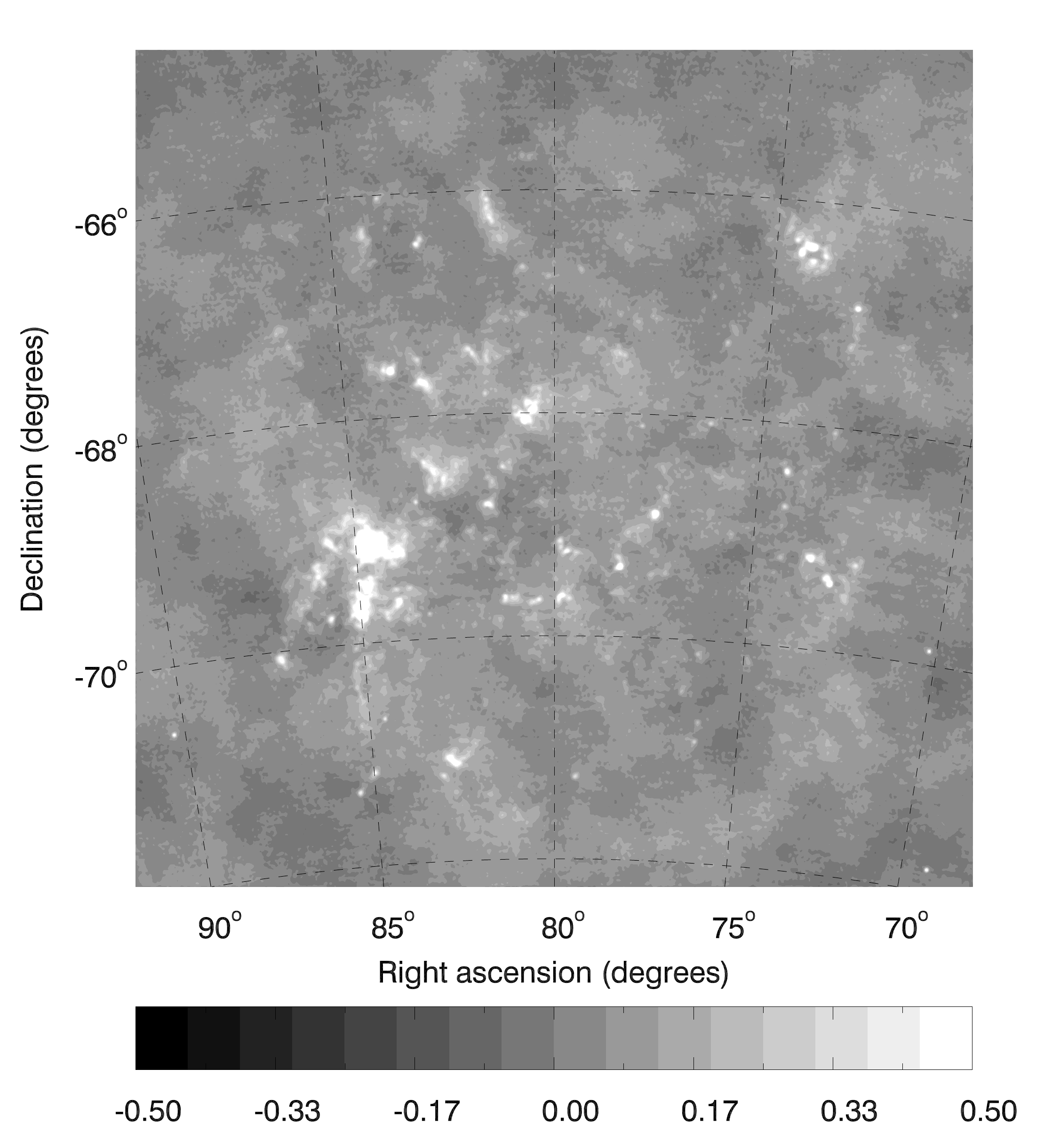}}
\subfigure[SMC, 500~$\mu$m, 2.0-arcmin resolution]{
  \includegraphics[width=3.25in]{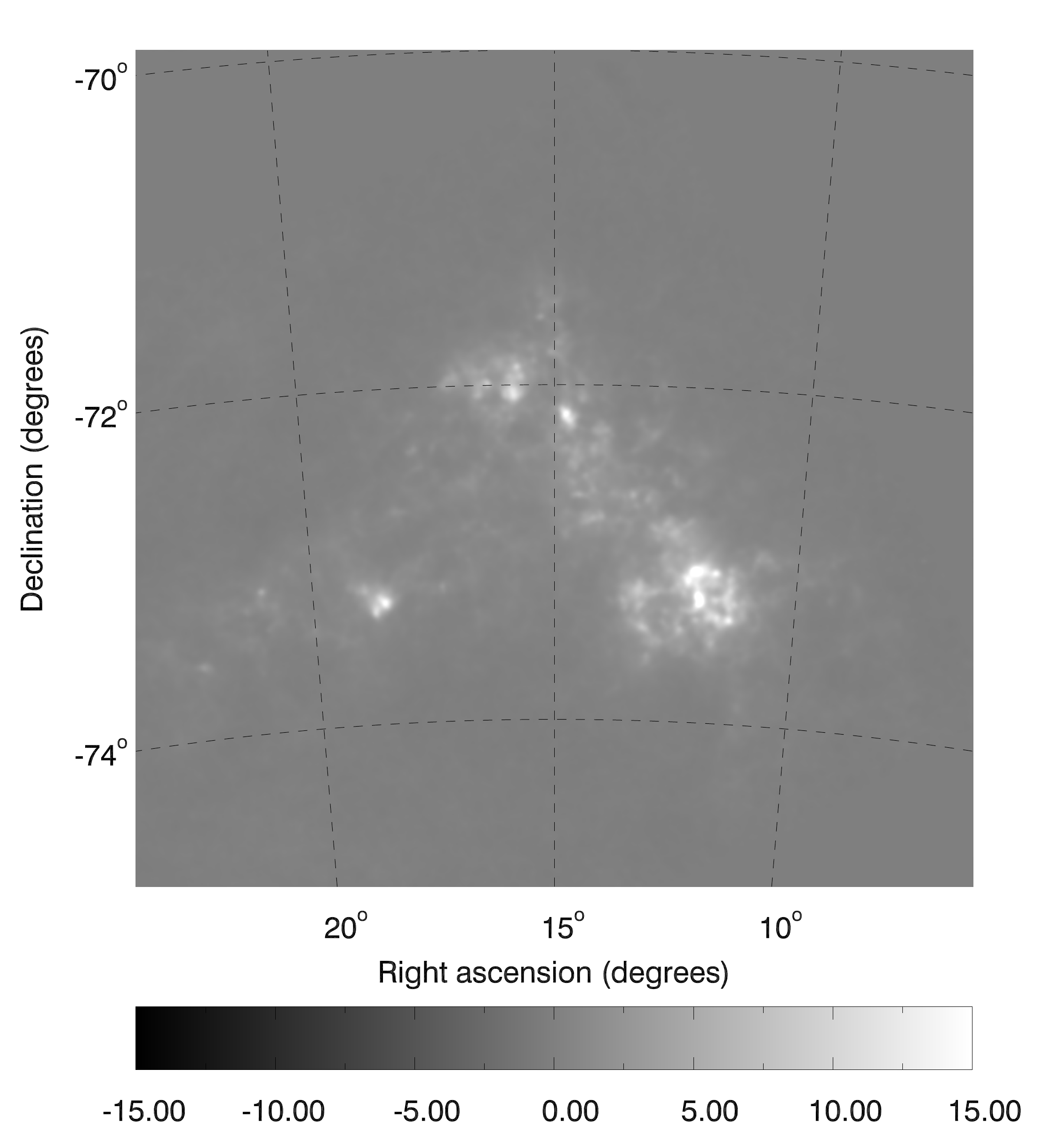}}
\subfigure[SMC, 2.1~mm, 2.0-arcmin resolution]{
  \includegraphics[width=3.25in]{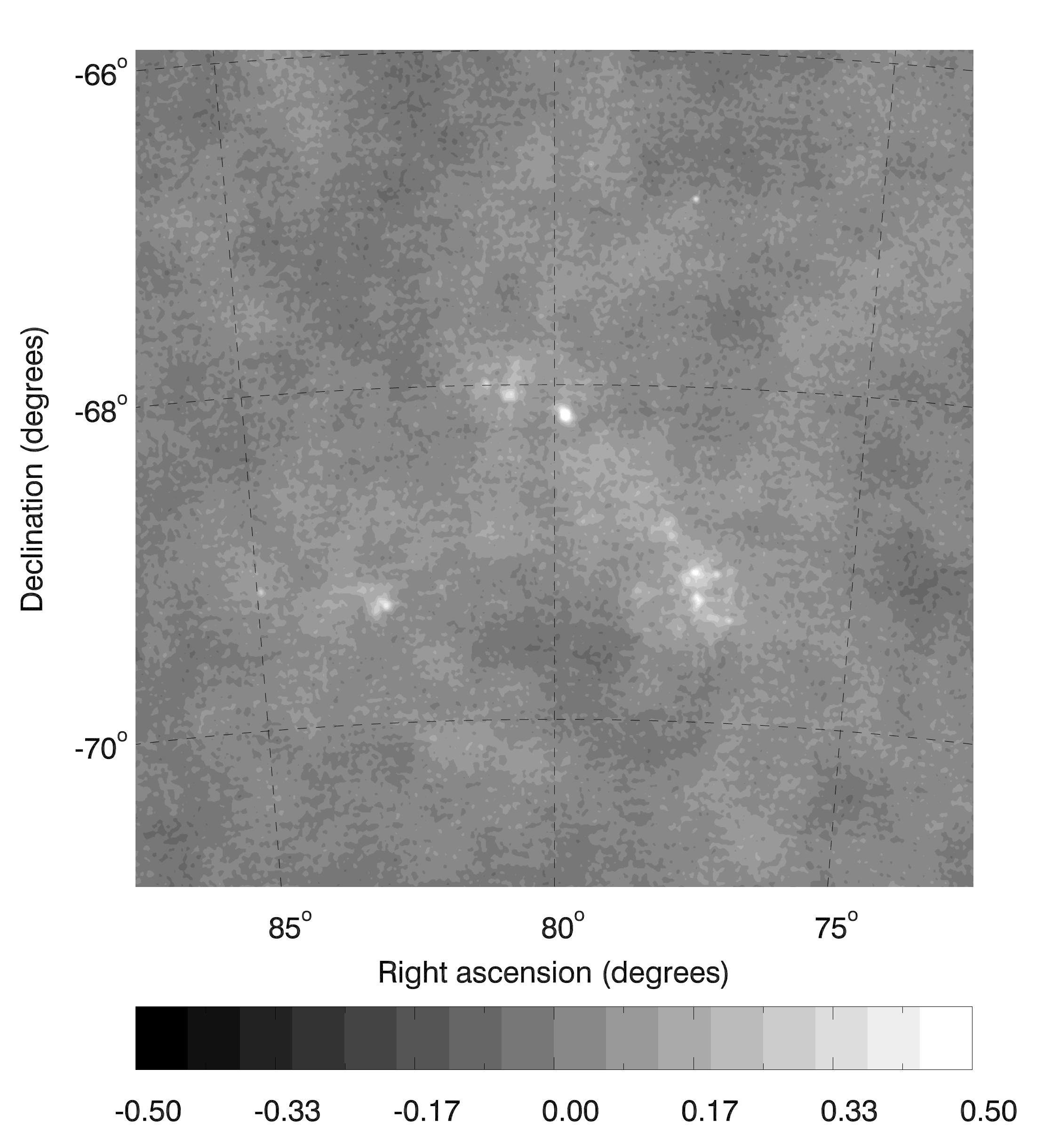}}
\caption{
\label{fig:lmcmapswspire} 
\herschel-SPIRE and combined SPT-\planck\ maps of the LMC and SMC.
{\bf Top row}: 500~$\mu$m \herschel-SPIRE map ({\it Left Panel})
and 2.1~mm combined SPT-\planck\ map ({\it Right Panel})
of the 7.5-by-7.5-degree LMC field at 2.0~arcmin resolution.
{\bf Bottom row}: 500~$\mu$m \herschel-SPIRE map ({\it Left Panel})
and 2.1~mm combined SPT-\planck\ map ({\it Right Panel})
of the 5-by-5-degree SMC field at 2.0~arcmin resolution.
The \herschel-SPIRE maps 
are the publicly available maps from \citet{meixner13} and
have been reprojected from their original projection and map center
to the same projection and map center used for the
combined SPT-\planck\ maps and smoothed with a Gaussian kernel with FWHM equal
to the quadrature difference of 2.0~arcmin and the SPIRE beam at each wavelength.
The combined SPT-\planck\ maps were constructed assuming 
an emission spectrum $I(\lambda) \propto \lambda^{-2}$.
The units of all maps are MJy~sr$^{-1}$.
}
\end{centering}
\end{figure*}

In Figure~\ref{fig:lmcmapswspire}, we show images of the LMC and SMC in two representative 
bands---the SPIRE 500~$\mu$m band and the SPT-\planck\ 2.1~mm band---at a common resolution.
To produce the SPIRE maps, we download the publicly available HERITAGE maps 
from the NASA/IPAC Infrared Science Archive\footnote{\url{http://irsa.ipac.caltech.edu/data/Herschel/HERITAGE/}},
reproject them from the native projection and map center to the projection and map center used
in this work, and convolve them with a Gaussian kernel with FWHM equal to the quadrature
difference between 2~arcmin and the SPIRE beam FWHM in each band. 

\begin{figure*}
\begin{center}
\includegraphics[width=7in]{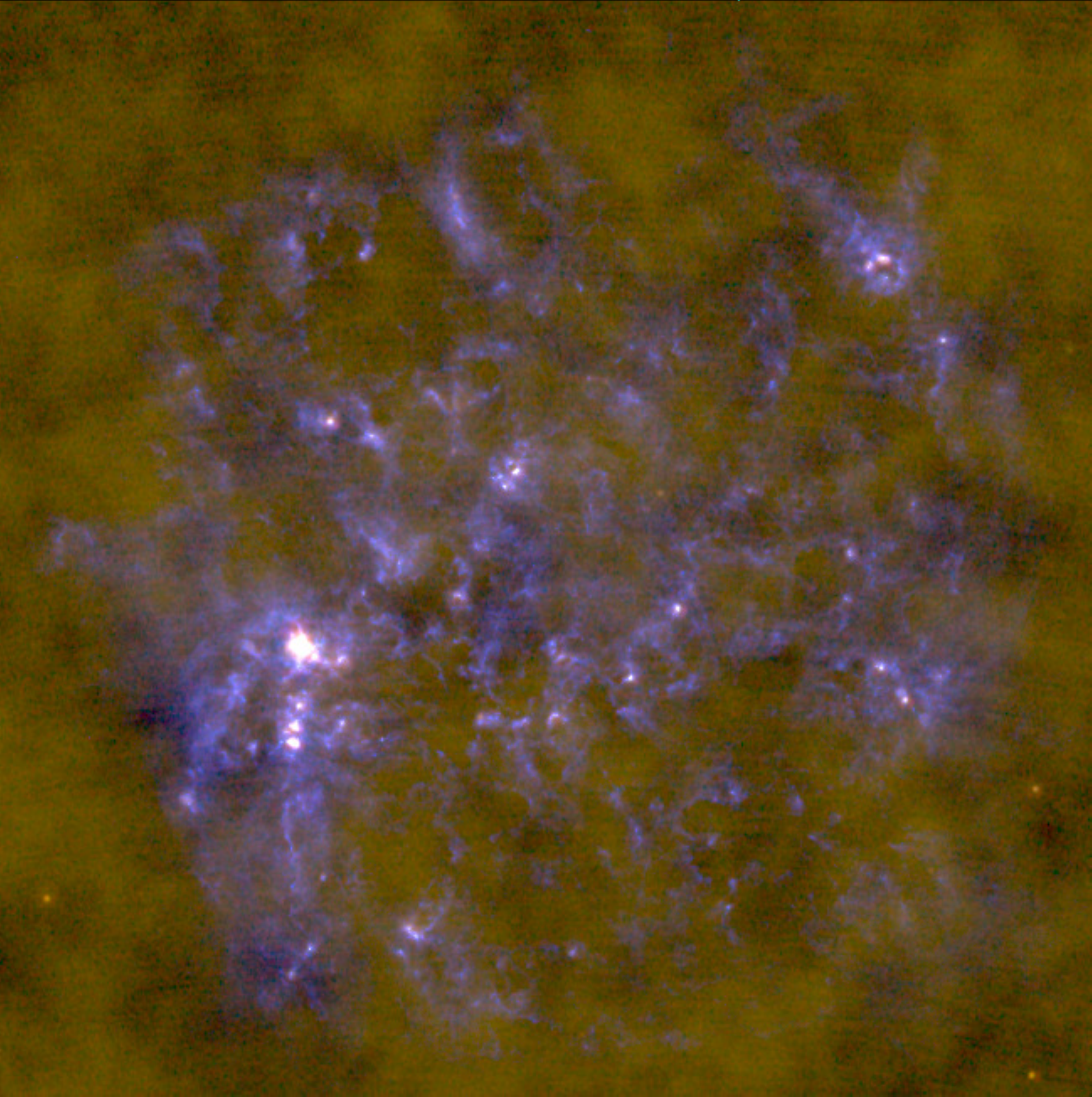}
\end{center}
\caption{
Three-color image of the LMC. Red = SPT-\planck\ 3.0~mm; 
Green = SPT-\planck\ 2.1~mm; Blue = 500~$\mu$m \herschel-SPIRE. 
The respective scales are $[-0.08,1.0]$, $[-0.08,2.5]$,
and $[-0.08,35.0]$~MJy~sr$^{-1}$,
such that a source with a $\lambda^{-2}$ spectrum would appear roughly white.
Before combining, all three maps have been convolved with a smoothing kernel such that the 
resolution in the map is 2.0~arcmin.
The diffuse, large-scale, yellowish signal is anisotropy in the CMB. Most of the filamentary structure
in the LMC is blue, indicating thermal dust as the primary emission mechanism. The bright
knots are in general redder, indicating a higher fraction of free-free or synchrotron emission, consistent with  
\planck-only results in \citet{planck11-17} and Section~\ref{sec:planckonly}. (Note that 
the very brightest regions such as 30-Doradus appear white because they saturate the color scale, not because
they are exactly consistent with $\lambda^{-2}$.) The red, point-like sources at the perimeter of the
image are background radio sources, all of which have counterparts in the SUMSS catalog \citep{mauch03}.
This image was produced using STIFF~\citep{bertin12}.
}
\label{fig:rgb2}
\end{figure*}

\begin{figure*}
\begin{center}
\includegraphics[width=7in]{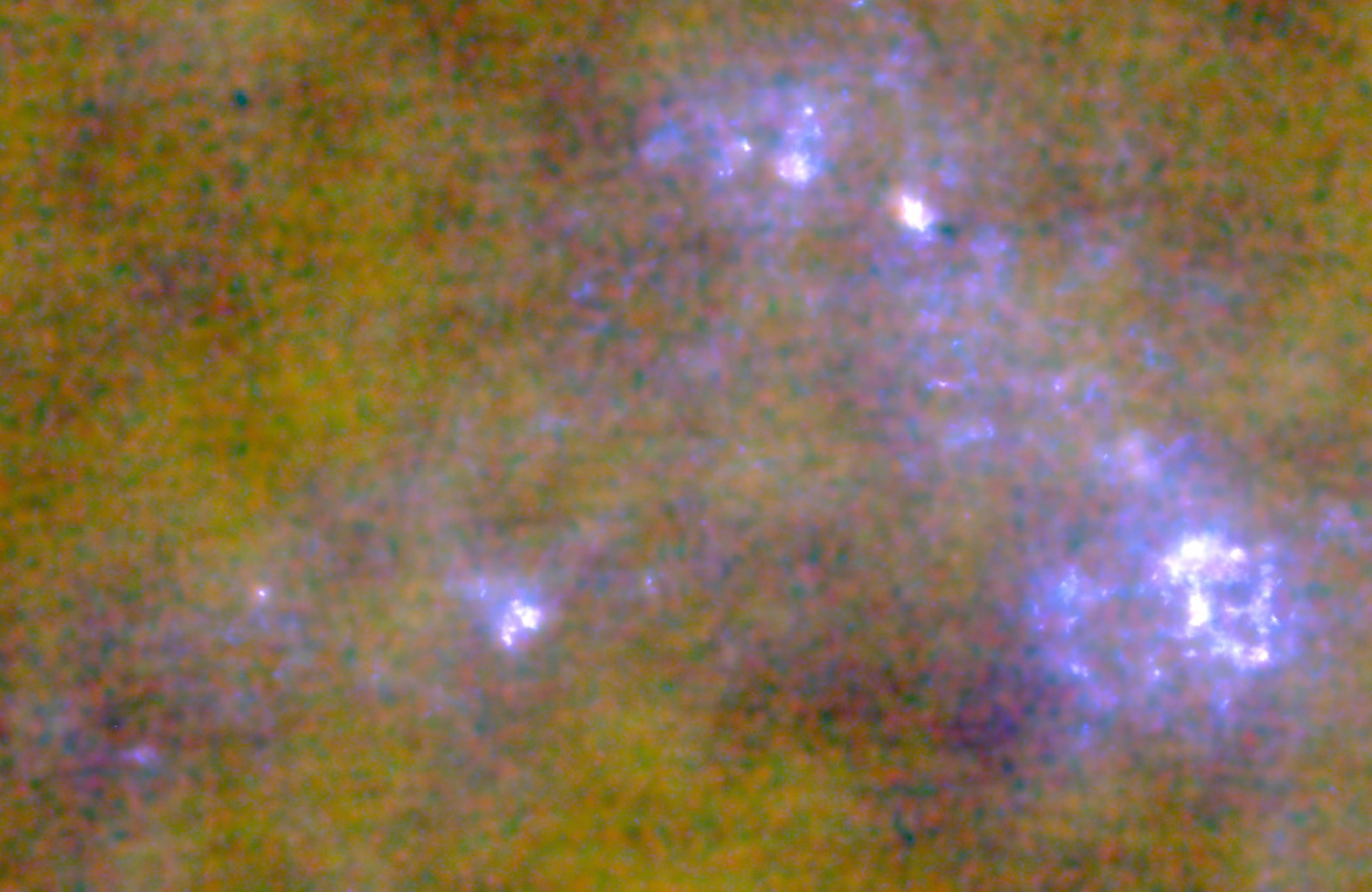}
\end{center}
\caption{
Three-color image of the SMC. Red = SPT-\planck\ 3.0~mm; 
Green = SPT-\planck\ 2.1~mm; Blue = 500~$\mu$m \herschel-SPIRE. 
The respective scales are $[-0.1,0.4]$, $[-0.1,0.8]$,
and $[-0.1,7.0]$~MJy~sr$^{-1}$,
such that a source with a $\lambda^{-2}$ spectrum would appear roughly white.
Before combining, all three maps have been convolved with a smoothing kernel such that the 
resolution in the map is 2.5~arcmin.
The diffuse, large-scale, yellowish signal is anisotropy in the CMB.
This image was produced using STIFF~\citep{bertin12}.
}
\label{fig:rgb3}
\end{figure*}

A high level of common structure is evident between the two bands shown in 
Figure~\ref{fig:lmcmapswspire}, but the densest, brightest knots of emission are more
prominent relative to the diffuse structure in the 2.1~mm map.
This is consistent with the indications from \planck-only data in 
\citet{planck11-17} and Section~\ref{sec:planckonly}
that the brightest regions have a higher contribution from synchrotron and free-free emission than
the filaments, particularly in the LMC.
The three-color images in Figures~\ref{fig:rgb2} and \ref{fig:rgb3} reinforce this picture. These images combine
resolution-matched SPT-\planck\ 3.0~mm (red) and 2.1~mm (green) and SPIRE 500~$\mu$m (blue)
maps (at 2.0~arcmin resolution for the LMC and 2.5~arcmin resolution for the SMC)
with a relative scaling such that emission that scales as $\lambda^{-2}$ would appear roughly white.
As expected, the filamentary structure of the LMC and SMC appears mostly blue, consistent with thermal 
dust emission going as $\lambda^{-\alpha}$ with $\alpha > 2$, while the dense, bright knots of 
emission are redder (where they do not saturate the color scale). The diffuse, yellowish background
in these images is the CMB, while the red, unresolved sources are background radio galaxies,
all of which have counterparts in the 36~cm (843~MHz) Sydney University Molongolo Sky Survey 
(SUMSS, \citealt{mauch03}).

\section{Conclusions}
\label{sec:conclusions}

We have created maps of the Large and Small Magellanic Clouds from combined SPT 
and \planck\ data in three wavelength bands, centered at roughly
1.4, 2.1, and 3.0~mm. These maps---one set of 40 maps each for the LMC and SMC
fields---consist of eight maps each created assuming different underlying emission
spectra (power-law emission $I(\lambda) \propto \lambda^{-\alpha}$ with different spectral indices).
We have created maps at three different final resolutions (Gaussian FWHM of 1.5, 2.0, and 2.5~arcmin)
in the 1.4 and 2.1~mm bands and two different resolutions (2.0 and 2.5~arcmin) in the 3.0~mm
band. For each set of maps assuming a given spectral index, we have calibrated and color-corrected
the SPT data to match the \planck\ data in a given band. We have then used knowledge of the
noise properties and angular response function for each map to make an inverse-variance-weighted
combination of the two instruments' data as a function of angular scale.

We have performed several consistency checks on the resulting maps, and we have 
estimated the noise contributions from instrumental and astrophysical components to 
flux measurements performed on those maps. We have visually compared the maps of the LMC
to FIR/submm maps from the \herschel\ HERITAGE survey and found clear common structure 
and evidence of a dependence of emission mechanism on brightness and/or density.

These maps extend the angular resolution of mm-wave studies of the Magellanic Clouds
down to $\sim$1~arcmin---or, equivalently, extend the wavelength coverage of arcminute-scale
maps of the Magellanic Clouds into the mm-wave regime. 
We expect these maps to be useful resources in studies of star formation in diverse 
environments and to increase our understanding of the physical processes at work in our
two nearest neighbor galaxies.
All data products described in this paper are available for download at 
\url{http://pole.uchicago.edu/public/data/maps/magclouds}
and from the NASA Legacy Archive for Microwave Background Data Analysis server.

\begin{acknowledgements}
The South Pole Telescope is supported by the National Science Foundation through grant PLR-1248097.  Partial support is also provided by the NSF Physics 
Frontier Center grant PHY-1125897 to the Kavli Institute of Cosmological Physics at the University of Chicago, the Kavli Foundation and the Gordon and 
Betty Moore Foundation grant GBMF 947. The McGill group acknowledges funding from the National Sciences and Engineering Research Council of Canada, Canada Research Chairs program, and the Canadian Institute for Advanced Research.
Argonne National Laboratory work was supported under U.S. Department of Energy contract DE-AC02-06CH11357.
We thank M. Meixner and the HERITAGE team for making their data publicly available and 
K. Ganga for helpful discussion on \planck\ map properties.
\end{acknowledgements}

{\it Facilities:}
\facility{Herschel},
\facility{Planck},
\facility{South Pole Telescope}

\clearpage

\begin{deluxetable*}{c c c l l l l l l}
\tabletypesize{\tiny}
\tablecaption{Noise contributions as a function of aperture diameter for the maps constructed 
assuming spectral index $\alpha=2.0$ and with 2.0~arcmin resolution.}
\tablehead{
\colhead{Field} &
\colhead{Aperture Radius} & 
\colhead{Compensated?} &
\multicolumn{3}{c}{Instrument Noise} & 
\multicolumn{3}{c}{CMB Noise} \\ 
\colhead{} & 
\colhead{[arcmin]} & 
\colhead{} & 
\multicolumn{3}{c}{[mJy]} & 
\multicolumn{3}{c}{[mJy]} \\ 
\colhead{} &
\colhead{} &
\colhead{} & 
\colhead{1.4~mm} & 
\colhead{2.1~mm} &
\colhead{3.2~mm} &
\colhead{1.4~mm} & 
\colhead{2.1~mm} &
\colhead{3.2~mm} 
}
\startdata
LMC &        2 & N &   20.5 &    5.4 &    9.6 &   46.9 &   38.7 &   24.3 \\
 &        3 & N &   17.8 &    8.1 &   12.5 &  104.9 &   86.7 &   54.4 \\
 &        4 & N &   27.9 &    9.6 &   14.9 &  187.3 &  154.8 &   97.2 \\
 &        5 & N &   28.8 &   11.1 &   16.3 &  289.1 &  238.9 &  150.1 \\
 &        7 & N &   37.9 &   14.1 &   19.0 &  547.1 &  452.1 &  284.0 \\
 &       10 & N &   46.8 &   18.4 &   24.4 & 1055.9 &  872.5 &  548.1 \\
 &       15 & N &   57.7 &   27.6 &   30.9 & 2160.9 & 1785.6 & 1121.7 \\
 &       20 & N &   75.1 &   34.4 &   40.6 & 3500.9 & 2893.0 & 1817.3 \\
 &       30 & N &   99.4 &   52.3 &   53.5 & 6547.0 & 5410.0 & 3398.4 \\
LMC &        2 & Y &   38.2 &    5.2 &   10.8 &    2.3 &    1.9 &    1.2 \\
 &        3 & Y &   45.5 &    9.9 &   19.0 &    9.5 &    7.8 &    4.9 \\
 &        4 & Y &   68.7 &   15.3 &   25.7 &   23.9 &   19.8 &   12.4 \\
 &        5 & Y &   83.8 &   18.3 &   32.1 &   48.3 &   39.9 &   25.1 \\
 &        7 & Y &   99.6 &   23.4 &   44.9 &  125.3 &  103.6 &   65.1 \\
 &       10 & Y &   82.5 &   32.9 &   57.6 &  317.1 &  262.0 &  164.6 \\
 &       15 & Y &  116.1 &   44.4 &   73.9 &  838.5 &  692.9 &  435.3 \\
 &       20 & Y &  156.7 &   50.1 &   77.6 & 1656.9 & 1369.2 &  860.1 \\
 &       30 & Y &  189.3 &   70.1 &  105.1 & 4158.5 & 3436.3 & 2158.6 \\
SMC &        2 & N &   32.9 &    8.0 &   15.2 &   46.9 &   38.7 &   24.3 \\
 &        3 & N &   28.9 &   11.3 &   18.9 &  104.9 &   86.7 &   54.4 \\
 &        4 & N &   46.8 &   12.0 &   22.1 &  187.3 &  154.8 &   97.2 \\
 &        5 & N &   44.8 &   12.6 &   24.5 &  289.1 &  238.9 &  150.1 \\
 &        7 & N &   59.0 &   16.0 &   27.2 &  547.1 &  452.1 &  284.0 \\
 &       10 & N &   70.6 &   20.1 &   36.2 & 1055.9 &  872.5 &  548.1 \\
 &       15 & N &   92.9 &   26.5 &   44.1 & 2160.9 & 1785.6 & 1121.7 \\
 &       20 & N &  111.8 &   33.4 &   50.8 & 3500.9 & 2893.0 & 1817.3 \\
 &       30 & N &  144.5 &   53.0 &   68.4 & 6547.0 & 5410.0 & 3398.4 \\
SMC &        2 & Y &   56.8 &    8.4 &   17.5 &    2.3 &    1.9 &    1.2 \\
 &        3 & Y &   68.2 &   16.9 &   31.2 &    9.5 &    7.8 &    4.9 \\
 &        4 & Y &  104.6 &   23.6 &   45.1 &   23.9 &   19.8 &   12.4 \\
 &        5 & Y &  112.1 &   26.5 &   53.1 &   48.3 &   39.9 &   25.1 \\
 &        7 & Y &  133.0 &   37.1 &   70.2 &  125.3 &  103.6 &   65.1 \\
 &       10 & Y &  113.4 &   47.4 &   84.2 &  317.1 &  262.0 &  164.6 \\
 &       15 & Y &  176.9 &   57.0 &  100.1 &  838.5 &  692.9 &  435.3 \\
 &       20 & Y &  214.9 &   59.5 &  108.0 & 1656.9 & 1369.2 &  860.1 \\
 &       30 & Y &  290.2 &   86.5 &  137.7 & 4158.5 & 3436.3 & 2158.6 \\
\enddata
\tablecomments{Noise levels for maps constructed using other assumed values of spectral
index are within 20\% of the values in this table. The noise levels for other map resolutions
are very similar for the CMB in all aperture sizes and for instrument noise at aperture
sizes larger than either map's resolution.}
\label{tab:noisevrad}
\end{deluxetable*}

\bibliography{../../../../BIBTEX/spt}
\end{document}